\documentclass[12pt,english]{article}
\usepackage{times}
\usepackage[T1]{fontenc}
\usepackage{geometry}
\usepackage{rotating}
\usepackage{graphicx}
\usepackage{setspace}
\usepackage{amssymb}

\makeatletter

\providecommand{\tabularnewline}{\\}

\usepackage{bm}

\usepackage{babel}
\makeatother
\begin{document}

\section*{\noindent One-Time Programmable Passive Electromagnetic Skins}

\noindent \vfill

\noindent G. Oliveri,$^{(1)(2)}$ \emph{Fellow}, \emph{IEEE}, F. Zardi,$^{(1)(2)}$
\emph{Member}, \emph{IEEE}, A. Salas-Sanchez,$^{(1)(2)}$ \emph{Member},
\emph{IEEE}, and A. Massa,$^{(1)(2)(3)(4)(5)}$ \emph{Fellow, IEEE}

\noindent \vfill

\noindent {\footnotesize $^{(1)}$} \emph{\footnotesize ELEDIA Research
Center} {\footnotesize (}\emph{\footnotesize ELEDIA}{\footnotesize @}\emph{\footnotesize UniTN}
{\footnotesize - University of Trento)}{\footnotesize \par}

\noindent {\footnotesize DICAM - Department of Civil, Environmental,
and Mechanical Engineering}{\footnotesize \par}

\noindent {\footnotesize Via Mesiano 77, 38123 Trento - Italy}{\footnotesize \par}

\noindent \textit{\emph{\footnotesize E-mail:}} {\footnotesize \{}\emph{\footnotesize giacomo.oliveri,
francesco.zardi, aaron.salassanchez@unitn.it, andrea.massa}{\footnotesize \}@}\emph{\footnotesize unitn.it}{\footnotesize \par}

\noindent {\footnotesize Website:} \emph{\footnotesize www.eledia.org/eledia-unitn}{\footnotesize \par}

\noindent {\footnotesize ~}{\footnotesize \par}

\noindent {\footnotesize $^{(2)}$} \emph{\footnotesize CNIT - \char`\"{}University
of Trento\char`\"{} ELEDIA Research Unit }{\footnotesize \par}

\noindent {\footnotesize Via Sommarive 9, 38123 Trento - Italy}{\footnotesize \par}

\noindent {\footnotesize Website:} \emph{\footnotesize www.eledia.org/eledia-unitn}{\footnotesize \par}

\noindent {\footnotesize ~}{\footnotesize \par}

\noindent {\footnotesize $^{(3)}$} \emph{\footnotesize ELEDIA Research
Center} {\footnotesize (}\emph{\footnotesize ELEDIA}{\footnotesize @}\emph{\footnotesize UESTC}
{\footnotesize - UESTC)}{\footnotesize \par}

\noindent {\footnotesize School of Electronic Science and Engineering,
Chengdu 611731 - China}{\footnotesize \par}

\noindent \textit{\emph{\footnotesize E-mail:}} \emph{\footnotesize andrea.massa@uestc.edu.cn}{\footnotesize \par}

\noindent {\footnotesize Website:} \emph{\footnotesize www.eledia.org/eledia}{\footnotesize -}\emph{\footnotesize uestc}{\footnotesize \par}

\noindent {\footnotesize ~}{\footnotesize \par}

\noindent {\footnotesize $^{(4)}$} \emph{\footnotesize ELEDIA Research
Center} {\footnotesize (}\emph{\footnotesize ELEDIA@TSINGHUA} {\footnotesize -
Tsinghua University)}{\footnotesize \par}

\noindent {\footnotesize 30 Shuangqing Rd, 100084 Haidian, Beijing
- China}{\footnotesize \par}

\noindent {\footnotesize E-mail:} \emph{\footnotesize andrea.massa@tsinghua.edu.cn}{\footnotesize \par}

\noindent {\footnotesize Website:} \emph{\footnotesize www.eledia.org/eledia-tsinghua}{\footnotesize \par}

\noindent {\small ~}{\small \par}

\noindent {\small $^{(5)}$} {\footnotesize School of Electrical Engineering}{\footnotesize \par}

\noindent {\footnotesize Tel Aviv University, Tel Aviv 69978 - Israel}{\footnotesize \par}

\noindent \textit{\emph{\footnotesize E-mail:}} \emph{\footnotesize andrea.massa@eng.tau.ac.il}{\footnotesize \par}

\noindent {\footnotesize Website:} \emph{\footnotesize https://engineering.tau.ac.il/}{\footnotesize \par}

\noindent \vfill

\noindent \emph{This work has been submitted to the IEEE for possible
publication. Copyright may be transferred without notice, after which
this version may no longer be accessible.}

\noindent \vfill

\newpage
\section*{One-Time Programmable Passive Electromagnetic Skins}

~

~

~

\begin{flushleft}G. Oliveri, F. Zardi, A. Salas-Sanchez, and A. Massa\end{flushleft}

\vfill

\begin{abstract}
\noindent The implementation of simple, inexpensive, and mass-production-oriented
solutions for smart electromagnetic environments (\emph{SEME}s) is
dealt with by introducing the concept of {}``one-time programmable''
electromagnetic skins (\emph{OTP-EMS}s). The simultaneous achievement
of modular fabrication, (one-time) configurable reflection properties,
passive-static operation, and zero maintenance is yielded by integrating
\emph{expendable components} at the atomic level of \emph{EMS}s. Towards
this end, an \emph{OTP} meta-atom structure is properly defined and
optimized to build \emph{EMS}s featuring the desired scenario-dependent
\emph{EM} wave manipulation functionalities. In order to illustrate
the features as well as to point out the potentialities of \emph{OTP-EMS}s,
a representative set of analytical, numerical, and experimental results
is reported by considering different apertures, illuminations, and
EM wave manipulation requirements.

\vfill
\end{abstract}
\noindent \textbf{Key words}: Reconfigurable Passive \emph{EM} Skins;
Static Passive \emph{EM} Skins; Smart Electromagnetic Environment;
Next-Generation Communications; Expendable EM Components.

\newpage
\section{Introduction and Rationale\label{sec:Introduction}}

\noindent Electromagnetic skins (\emph{EMS}s) are one of the most
important technologies in Smart Electromagnetic Environments (\emph{SEME}s)
\cite{Massa 2021}\cite{Oliveri 2021c}. \emph{EMS}s, which are often
implemented as two dimensional metamaterials with suitably-designed
local reflection properties \cite{Oliveri 2015}\cite{Yang 2022},
are able to tailor the wireless propagation in both outdoor and indoor
scenarios \cite{Yang 2022}\cite{Oliveri 2024}. Thanks to this latter
property, there has been a considerable interest, both in academia
and industry, to their development and successive integration within
next-generation communication and sensing systems \cite{Yang 2022}.
Therefore, different classes of \emph{EMS}s have been introduced \cite{Massa 2021}\cite{Oliveri 2021c}\cite{Yang 2022}\cite{Oliveri 2024}-\cite{Alu 2024}. 

\noindent Reconfigurable passive \emph{EMS}s (\emph{RP-EMS}s), also
known as reconfigurable intelligent surfaces (\emph{RIS}s), have been
widely studied for enhancing the network performance \cite{Di Renzo 2019}-\cite{Zardi 2024}.
However, current implementation of \emph{RIS}s are often characterized
by non-trivial fabrication/operation costs regardless of the adopted
technology \cite{Oliveri 2015} and they may imply a non-negligible
system overhead owing to the increased complexity in the channel sensing
and network control \cite{Zardi 2024}. Otherwise, static passive
\emph{EMS}s (\emph{SP-EMS}s) are an inexpensive and zero-consumption
zero-overhead solution \cite{Oliveri 2021c}\cite{Oliveri 2024} for
the control of the \emph{EM} propagation in \emph{SEME}s. Unfortunately,
since the working mechanism of \emph{SP-EMS}s is based on the local
geometrical/physical properties of their single meta-atoms \cite{Oliveri 2021c},
dedicated designs as well as (successive) fabrication processes are
needed for each functionality/scenario at hand, hence preventing mass
production.

\noindent Starting from these considerations, an alternative technological
solution for the \emph{EM} propagation control is proposed in the
following to simultaneously enable (\emph{i}) simple, inexpensive,
and mass production-oriented fabrication; (\emph{ii}) modular deployment,
but one-time configurable performance, to customize the wave control
functionality to the scenario of interest; (\emph{iii}) passive-static
structure and no-maintenance operation once installed. The basic idea
to fulfill these contrasting requirements is that of integrating at
least one \emph{expendable} component (e.g., an electrical fuse {[}Fig.
1(\emph{b}){]}) into the static passive meta-atom structure {[}Fig.
1(\emph{a}){]} so that its reflection behaviour depends on the integrity
of the component itself. The \emph{EMS} obtained by the combination
of this class of unit cells {[}Fig. 2(\emph{a}){]} turns out to be
{}``one-time programmable'' (\emph{OTP-EMS}) during installation,
while it works as a standard \emph{SP-EMS} during operation. As a
matter of fact, the user can set the local reflection properties of
each atom by simply applying a sufficiently large current to its \emph{expendable}
component {[}Fig. 2(\emph{b}){]}. Thus, the fabrication process turns
out to be fully modular and mass production-oriented, while the customization
of such an \emph{EMS} to the user-required \emph{EM} wave manipulation
functionality is yielded with a suitable configuration step when installed
in the scenario at hand.

\noindent Despite the promising features and potentialities, several
challenges need to be addressed to demonstrate the feasibility as
well as the efficiency of \emph{OTP-EMS}s. Indeed, an \emph{ad-hoc}
meta-atom must be designed so that, when the expendable component
is burnt, a suitable (i.e., sufficiently large) phase change in the
reflected \emph{EM} wave may be observed. Moreover, the impact of
the limited programmability (i.e., a reduced number of degrees-of-freedom)
of such an \emph{EMS} and the effects of the non-idealities of the
expendable components (e.g., the spurious resistance/inductance/capacitance
of the fuse) on the resulting \emph{EM} field manipulation properties
have to be carefully taken into account. The objective of this work
is to address such challenges and to evaluate, both numerically and
experimentally, the effectiveness of the synthesized \emph{OTP-EMS}s
in affording realistic phase manipulation functionalities.

\noindent To the best of the authors' knowledge, the main innovative
contributions of this research work include (\emph{i}) the introduction
within the \emph{SEME} framework of the concept of \emph{OTP-EMS}s
as a complement to existing SoA static/reconfigurable \emph{EMS} architectures,
(\emph{ii}) the proof that meta-atoms with expendable components can
be profitably employed as elementary building blocks of \emph{EMS}s
that afford one-time reconfigurable \emph{EM} wave manipulation features,
(\emph{iii}) the numerical and experimental assessment of the reliability
and the effectiveness of \emph{OTP-EMS}s built with commercially-available
fuses compliant with mass production surface mounting processes.

\noindent The outline of the paper is as follows. After the formulation
of the design problem at hand, the procedure for the synthesis of
both the meta-atom and the corresponding \emph{OTP-EMS} is presented
(Sect. \ref{sec:Problem-Formulation}). Representative results from
a set of numerical and experimental tests, concerned with both the
unit cell design (Sect. \ref{sub:ODR-Meta-Atom-Design}) and the \emph{OTP-EMS}
panel (Sect. \ref{sub:ODR-EMSs-Design-and}), are then reported to
give the interested readers some insights on the feasibility as well
as the reliability of the \emph{OPT-EMS} concept in different operative
conditions. Finally, some conclusions are drawn (Sect. \ref{sec:Conclusions-and-Remarks}).

\section{\noindent Problem Formulation and Design Process\label{sec:Problem-Formulation} }

\noindent Let us consider a standard \emph{SEME} scenario \cite{Massa 2021}\cite{Oliveri 2021c}
where an \emph{OTP-EMS}, composed of $P\times Q$-meta-atoms {[}Fig.
2(\emph{a}){]} and covering an area/aperture $\Omega$, is illuminated
by a time-harmonic plane wave with electric field $\mathbf{E}_{inc}$
and generated by a far-field source. Moreover, let $\mathbf{r}=\left(r,\theta,\varphi\right)$
be the position vector in the global coordinate system centered at
the \emph{OTP-EMS} barycenter.

\noindent According to the \emph{EMS} radiation theory \cite{Yang 2019}\cite{Oliveri 2021c},
the power pattern of the field reflected by the \emph{EMS} in the
Fraunhofer region is given by \cite{Balanis 1989}\begin{equation}
\begin{array}{c}
\mathcal{F}\left(\theta,\varphi;\underline{g},\underline{s}\right)=\frac{k^{2}}{2\eta}\left\{ \left|\left[\mathcal{N}_{\varphi}^{m}\left(\theta,\varphi;\underline{g},\underline{s}\right)+\eta\mathcal{N}_{\theta}^{e}\left(\theta,\varphi;\underline{g},\underline{s}\right)\right]\right|^{2}\right.+\\
+\left.\left|\left[\mathcal{N}_{\theta}^{m}\left(\theta,\varphi;\underline{g},\underline{s}\right)-\eta\mathcal{N}_{\varphi}^{e}\left(\theta,\varphi;\underline{g},\underline{s}\right)\right]\right|^{2}\right\} \end{array}\label{eq:power pattern}\end{equation}
where $k$ and $\eta$ are the free-space wavenumber and the free-space
impedance, respectively, while $\underline{g}$ is the vector of the
$D$ geometric/physical descriptors of the reference \emph{EMS} meta-atom
structure ($\underline{g}=\left\{ g^{\left(d\right)};\, d=1,...,D\right\} $).
Moreover, $\underline{s}=\left\{ s_{pq};p=1,...,P;\, q=1,...,Q\right\} $
is a vector whose $pq$-th ($p=1,...,P$; $q=1,...,Q$) entry $s_{pq}$
indicates the status of the \emph{expendable} component of the $pq$-th
meta-atom (i.e., $s_{pq}=0$ when the electrical fuse is burnt, and
$s_{pq}=1$ otherwise). Furthermore, $\mathcal{N}_{\alpha}^{w}$ is
the $\alpha$-th ($\alpha\in\left\{ \theta,\varphi\right\} $) spherical
component of the $w$-th ($w\in\left\{ e,m\right\} $; $e\to$electric,
$m\to$magnetic) radiation vector. In particular \cite{Yang 2019}\cite{Oliveri 2021c}\cite{Balanis 1989},\begin{equation}
\mathcal{N}_{\theta}^{w}\left(\theta,\varphi;\underline{g},\underline{s}\right)=\int_{\Omega}\left[J_{x}^{w}\left(\mathbf{r}';\underline{g},\underline{s}\right)\cos\theta\cos\varphi+J_{y}^{w}\left(\mathbf{r}';\underline{g},\underline{s}\right)\cos\theta\sin\varphi\right]\exp\left(jk\widehat{\mathbf{r}}\cdot\mathbf{r}'\right)\mathrm{d}\mathbf{r}'\label{eq:radiation theta}\end{equation}
\begin{equation}
\mathcal{N}_{\varphi}^{w}\left(\theta,\varphi;\underline{g},\underline{s}\right)=\int_{\Omega}\left[-J_{x}^{w}\left(\mathbf{r}';\underline{g},\underline{s}\right)\sin\varphi+J_{y}^{w}\left(\mathbf{r}';\underline{g},\underline{s}\right)\cos\varphi\right]\exp\left(jk\widehat{\mathbf{r}}\cdot\mathbf{r}'\right)\mathrm{d}\mathbf{r}'\label{eq:radiation phi}\end{equation}
where $\widehat{\mathbf{r}}=\left\{ \sin\theta\cos\varphi,\sin\theta\sin\varphi,\cos\varphi\right\} $
and $\mathbf{J}^{w}$ {[}$\mathbf{J}^{w}\left(\mathbf{r};\underline{g},\underline{s}\right)=\sum_{a=x,y}J_{a}^{w}\left(\mathbf{r};\underline{g},\underline{s}\right)\widehat{\mathbf{a}}${]}
is the equivalent $w$-th ($w\in\left\{ e,m\right\} $) surface ($\mathbf{r}\in\Omega$)
current whose expression, according to the Love's equivalence principle
and under the local periodicity approximation (\emph{LPA}) \cite{Yang 2019}\cite{Oliveri 2021c}\cite{Cuesta 2018},
turns out to be\begin{equation}
\mathbf{J}^{e}\left(\mathbf{r};\underline{g},\underline{s}\right)=\frac{1}{\eta}\widehat{\mathbf{z}}\times\mathbf{k}_{inc}\times\sum_{p=1}^{P}\sum_{q=1}^{Q}\overline{\overline{\Gamma}}\left(\underline{g};s_{pq};\mathbf{k}_{inc}\right)\cdot\mathbf{E}_{inc}\left(\mathbf{r}_{pq}\right)\Pi_{pq}\left(\mathbf{r}\right)\label{eq:correnti elettriche}\end{equation}
and

\noindent \begin{equation}
\mathbf{J}^{m}\left(\mathbf{r};\underline{g},\underline{s}\right)=-\widehat{\mathbf{z}}\times\sum_{p=1}^{P}\sum_{q=1}^{Q}\overline{\overline{\Gamma}}\left(\underline{g};s_{pq};\mathbf{k}_{inc}\right)\cdot\mathbf{E}_{inc}\left(\mathbf{r}_{pq}\right)\Pi_{pq}\left(\mathbf{r}\right),\label{eq:correnti}\end{equation}
$\mathbf{r}_{pq}$ {[}$\mathbf{r}_{pq}=\left(x_{p},y_{q},0\right)${]}
being the barycenter of the $pq$-th ($p=1,...,P$; $q=1,...,Q$)
cell with area $\Omega_{pq}$ ($\to$ $\Omega=\sum_{p=1}^{P}\sum_{q=1}^{Q}\Omega_{pq}$),
while $\mathbf{k}_{inc}$ is the incident wave vector in correspondence
with the incidence angle $\left(\theta_{inc},\varphi_{inc}\right)$,
$\Pi_{pq}\left(\mathbf{r}\right)$ is the $pq$-th ($p=1,...,P$;
$q=1,...,Q$) pixel basis function {[}$\Pi_{pq}\left(\mathbf{r}\right)=1$
if $\mathbf{r}\in\Omega_{pq}$, $\Pi_{pq}\left(\mathbf{r}\right)=0$
otherwise), and\begin{equation}
\overline{\overline{\Gamma}}\left(\underline{g};s_{pq};\mathbf{k}_{inc}\right)=\left[\begin{array}{cc}
\Gamma^{TE}\left(\underline{g};s_{pq};\mathbf{k}_{inc}\right) & \Gamma^{TE-TM}\left(\underline{g};s_{pq};\mathbf{k}_{inc}\right)\\
\Gamma^{TM-TE}\left(\underline{g};s_{pq};\mathbf{k}_{inc}\right) & \Gamma^{TM}\left(\underline{g};s_{pq};\mathbf{k}_{inc}\right)\end{array}\right]\label{eq:Gamma local}\end{equation}
is the local reflection tensor of the $pq$-th ($p=1,...,P$; $q=1,...,Q$)
meta atom.

\noindent It is worth remarking that, according to the adopted \emph{LPA},
(\ref{eq:power pattern}) neglects the edge and the \emph{a-periodic}
coupling effects taking place in the \emph{EMS} \cite{Yang 2019}\cite{Oliveri 2021c}.
Nevertheless, it has been proven accurate and reliable, both numerically
and experimentally, in most practical \emph{SEME} scenarios \cite{Oliveri 2021c}\cite{Oliveri 2022}\cite{Oliveri 2022b}-\cite{Oliveri 2023c}.
On the other hand, the previous derivation points out that here the
\emph{EMS} layout consists of geometrically identical meta-atoms,
since $\underline{g}$ does not depend on the unit cell $pq$-indexes,
and its {}``one-time programmable'' feature is yielded by acting
on the \emph{expendable} components embedded at the atomic level (i.e.,
$\underline{s}$) to control the corresponding reflection tensors
\{$\overline{\overline{\Gamma}}\left(\underline{g};s_{pq};\mathbf{k}_{inc}\right)$;
$p=1,...,P$; $q=1,...,Q$\}. Those characteristics fulfill all the
requirements for a \emph{}modular mass production-oriented fabrication
of \emph{OTP-EMS}s. Towards this end, the following two distinct problems
need to be properly addressed:

\begin{itemize}
\item \noindent \emph{Meta-Atom Design} (\emph{MAD}) \emph{Problem} - Identification
of a meta-atom architecture, $\underline{g}^{opt}$, that enables
a successive and proper control of the value of $\overline{\overline{\Gamma}}\left(\underline{g};s_{pq};\mathbf{k}_{inc}\right)$
by acting on $s_{pq}$;
\item \noindent \emph{OTP-EMS Synthesis} \emph{Problem} - Starting from
the meta-atom structure described by $\underline{g}^{opt}$, definition
of the optimal setup of the \emph{expendable} components, $\underline{s}^{opt}$,
to obtain the desired wave manipulation functionality.
\end{itemize}

\subsection{\noindent Meta-Atom Design \emph{}(\emph{MAD})\label{sub:Meta-Atom-Design-(MAD)}}

\noindent The \emph{MAD} problem can be mathematically stated as follows

\begin{quotation}
\noindent \emph{MAD Problem} - Given $\mathbf{k}_{inc}$, find the
optimal setup of the meta-atom geometrical descriptors, $\underline{g}^{opt}$,
that minimizes the \emph{OTP Meta-Atom} cost function $\Phi_{MAD}\left(\underline{g}\right)$
defined as\begin{equation}
\begin{array}{r}
\Phi_{MAD}\left(\underline{g}\right)=\beta_{1}\sum_{\zeta=TE,TM}\left|\Delta\Gamma^{\zeta}\left(\underline{g};\mathbf{k}_{inc}\right)-\pi\right|+\\
+\beta_{2}\sum_{\zeta=TE,TM}\sum_{s=0,1}\frac{1}{\left|\Gamma^{\zeta}\left(\underline{g};s;\mathbf{k}_{inc}\right)\right|}\end{array}\label{eq:design cell cost function}\end{equation}
(i.e., $\underline{g}^{opt}=\arg\left\{ \min_{\underline{g}}\left[\Phi_{MAD}\left(\underline{g}\right)\right]\right\} $).
\end{quotation}
\noindent In (\ref{eq:design cell cost function}), $\beta_{1}$ and
$\beta_{2}$ are user-defined real coefficients, while\begin{equation}
\Delta\Gamma^{\zeta}\left(\underline{g};\mathbf{k}_{inc}\right)=\left|\angle\Gamma^{\zeta}\left(\underline{g};s;\mathbf{k}_{inc}\right)_{s=1}-\angle\Gamma^{\zeta}\left(\underline{g};s;\mathbf{k}_{inc}\right)_{s=0}\right|\label{eq:}\end{equation}
 is the phase difference on the $\zeta$ ($\zeta\in\left\{ TE,TM\right\} $)
component of the local reflection tensor that the meta-atom exhibits
when the expendable component status changes from {}``normal'' ($s=1$)
to {}``burnt'' ($s=0$). The second additive term in (\ref{eq:design cell cost function})
is aimed at maximizing the magnitude of each $\zeta$ ($\zeta\in\left\{ TE,TM\right\} $)
component of the reflection coefficient in order to optimize the reflection
power efficiency of the \emph{EMS}.

\noindent Such a formulation of the {}``\emph{MAD Problem}'' turns
out to be a $D$-sized (the number of descriptors $D$ depending on
the unit cell complexity) real-variable non-linear optimization that,
according to optimization theory, can be efficiently solved by means
of an iterative evolutionary strategy \cite{Rocca 2009w}. More in
detail, a set of $C$ guess solutions, \{$\underline{g}_{c}^{\left(t\right)}$;
$c=1,...,C$\} ($C$ being the population size) are iteratively ($t$
being the iteration index) updated by undergoing the \emph{particle
swarm} operators \cite{Rocca 2009w} until either the maximum number
of iterations is reached (i.e., $t=T$) or a stagnation condition
on the cost function holds true \cite{Rocca 2009w}. The optimal unit
cell is finally outputted as $\underline{g}^{opt}=\arg\left\{ \min_{t=1,...,T}\left(\min_{c=1,...,C}\left[\Phi\left(\underline{g}_{c}^{\left(t\right)}\right)\right]\right)\right\} $.

\subsection{\noindent \emph{OTP-EMS} Synthesis\label{sub:OTP-EMS-Synthesis}}

\noindent Once the meta-atom has been synthesized by determining $\underline{g}^{opt}$,
the layout of the \emph{OTP-EMS} is physically implemented by replicating
$P\times Q$ geometrically-identical unit cells and one-time setting,
before the installation of the \emph{OTP-EMS} in the target location,
the user-desired functionality by optimizing the status of the arrangement
of $P\times Q$ \emph{expendable} components, $\underline{s}^{opt}$.

\noindent Of course, the optimal setup of the \emph{OTP-EMS} status
$\underline{s}^{opt}$ depends on the required wave manipulation function.
Without loss of generality, let us consider hereinafter the most commonly
required functionality \cite{Oliveri 2023c}\cite{Oliveri 2024},
which is the generation of a collimated reflection in an anomalous
angular direction $\left(\theta_{refl},\varphi_{refl}\right)$, and
let us formulate the {}``\emph{OTP-EMS Synthesis} \emph{Problem}''
as follows:

\begin{quotation}
\noindent \emph{OTP-EMS Synthesis} \emph{Problem} - By assuming the
meta-atom geometry coded in $\underline{g}^{opt}$ and given the desired
reflection direction $\left(\theta_{refl},\varphi_{refl}\right)$,
find the optimal setup of the \emph{EMS} status vector, $\underline{s}^{opt}$,
that minimizes the \emph{OTP-EMS} cost function $\Phi_{OTP-EMS}\left(\underline{g},\underline{s}\right)$
defined as \begin{equation}
\Phi_{OTP-EMS}\left(\underline{g},\underline{s}\right)=\frac{1}{\mathcal{F}\left(\theta_{refl},\varphi_{refl};\underline{g},\underline{s}\right)}\label{eq:cost function EMS}\end{equation}
 (i.e., $\underline{s}^{opt}=\arg\left\{ \min_{\underline{s}}\left[\Phi_{OTP-EMS}\left(\underline{g}^{opt},\underline{s}\right)\right]\right\} $).
\end{quotation}
\noindent A closed-form solution for such a problem can be derived
with the phase conjugation approach \cite{Yang 2019}\cite{Oliveri 2023b}\cite{Oliveri 2023c}
as detailed in the following.

\noindent For the sake of notation simplicity, let us refer to the
case of a \emph{TM}-polarized incident wave (i.e., $\mathbf{E}_{inc}\left(\mathbf{r}\right)=E_{inc}^{TM}\left(\mathbf{r}\right)\widehat{\psi}^{TM}$,
$\widehat{\psi}^{TM}$ being the unit vector for the \emph{TM} field
component) and let us assume negligible the depolarization effects
{[}$\to$ $\Gamma^{TM-TE}\left(\underline{g};s_{pq};\mathbf{k}_{inc}\right)=\Gamma^{TE-TM}\left(\underline{g};s_{pq};\mathbf{k}_{inc}\right)=0${]}.
Under these conditions, the expression of $\mathcal{N}_{\theta}^{e}\left(\theta,\varphi;\underline{s}\right)$
in (\ref{eq:power pattern}) simplifies into\begin{equation}
\mathcal{N}_{\theta}^{e}\left(\theta,\varphi;\underline{g},\underline{s}\right)=\int_{\Omega}\left[J_{x}^{e}\left(\mathbf{r}';\underline{g},\underline{s}\right)\cos\theta\cos\varphi+J_{y}^{e}\left(\mathbf{r}';\underline{g},\underline{s}\right)\cos\theta\sin\varphi\right]\exp\left(jk\widehat{\mathbf{r}}\cdot\mathbf{r}'\right)\mathrm{d}\mathbf{r}'\label{eq:vettore theta}\end{equation}
where $\mathbf{J}^{e}\left(\mathbf{r};\underline{g},\underline{s}\right)=\frac{1}{\eta}\widehat{\mathbf{z}}\times\mathbf{k}_{inc}\times\sum_{p=1}^{P}\sum_{q=1}^{Q}E_{inc}^{TM}\left(\mathbf{r}_{pq}\right)\Gamma^{TM}\left(\underline{g};s_{pq};\mathbf{k}_{inc}\right)\Pi_{pq}\left(\mathbf{r}\right)\widehat{\psi}^{TM}$.
By computing the integrals in (\ref{eq:vettore theta})\cite{Oliveri 2023b},
the expression of $\mathcal{N}_{\theta}^{e}$ reduces to

\noindent \begin{equation}
\begin{array}{c}
\mathcal{N}_{\theta}^{e}\left(\theta,\varphi;\underline{g},\underline{s}\right)=\frac{\widehat{\mathbf{z}}\cdot\mathbf{k}_{inc}}{\eta}\mathrm{sinc}\left(\frac{k\Delta_{x}\sin\theta\cos\varphi}{2}\right)\mathrm{sinc}\left(\frac{k\Delta_{y}\sin\theta\sin\varphi}{2}\right)\\
\left[\widehat{\psi}^{TM}\cdot\widehat{\mathbf{x}}\cos\theta\cos\varphi+\widehat{\psi}^{TM}\cdot\widehat{\mathbf{y}}\cos\theta\sin\varphi\right]\left[\widehat{\psi}^{TM}\cdot\widehat{\mathbf{x}}\cos\theta\cos\varphi+\widehat{\psi}^{TM}\cdot\widehat{\mathbf{y}}\cos\theta\sin\varphi\right]\\
\sum_{p=1}^{P}\sum_{q=1}^{Q}\Gamma^{TM}\left(\underline{g};s_{pq};\mathbf{k}_{inc}\right)E_{inc}^{TM}\left(\mathbf{r}_{pq}\right)\exp\left[jk\left(x_{p}\sin\theta\cos\varphi+y_{q}\sin\theta\sin\varphi\right)\right]\end{array}\label{eq:radiation vector simplified}\end{equation}
where $\mathrm{sinc}\left(\cdot\right)=\frac{\mathrm{\sin}\left(\cdot\right)}{\left(\cdot\right)}$,
while $\Delta_{x}$ and $\Delta_{y}$ are the unit cell periodicity
in $x$ and $y$, respectively. Analogous expressions can be derived
for the other components $\mathcal{N}_{\varphi}^{e}$, $\mathcal{N}_{\theta}^{m}$,
and $\mathcal{N}_{\varphi}^{m}$ in (\ref{eq:power pattern}).

\noindent By substituting (\ref{eq:radiation vector simplified})
e the analogous ones in (\ref{eq:power pattern}), it turns out that\begin{equation}
\mathcal{F}\left(\theta,\varphi;\underline{g},\underline{s}\right)=\frac{k^{2}}{2}\mathcal{A}\left(\theta,\varphi\right)\times\mathcal{P}\left(\theta,\varphi;\underline{g},\underline{s}\right)\label{eq:final power}\end{equation}
where $\mathcal{P}\left(\theta,\varphi;\underline{g},\underline{s}\right)$
$=$ $\left|\sum_{p=1}^{P}\right.\sum_{q=1}^{Q}\Gamma^{TM}\left(\underline{g};c_{pq};\mathbf{k}_{inc}\right)$
$E_{inc}^{TM}\left(\mathbf{r}_{pq}\right)$ $\exp$ {[}$jk$ $\left(x_{p}\sin\theta\cos\varphi\right.$
$+$ $\left.\left.\left.y_{q}\sin\theta\sin\varphi\right)\right]\right|^{2}$
and $\mathcal{A}\left(\theta,\varphi\right)$ $=$ $\left|\left(\widehat{\mathbf{z}}\cdot\mathbf{k}_{inc}\right)\right.$
$\mathrm{sinc}\left(\frac{k\Delta_{x}\sin\theta\cos\varphi}{2}\right)$
$\mathrm{sinc}\left(\frac{k\Delta_{y}\sin\theta\sin\varphi}{2}\right)$
$\left[\widehat{\psi}^{TM}\right.$ $\cdot$ $\left(\widehat{\mathbf{x}}\right.$
$\cos\theta\cos\varphi$ $+$ $\widehat{\mathbf{y}}\cos\theta$ $\left.\left.\left.\sin\varphi\right)\right]\right|^{2}$.

\noindent According to such a result (\ref{eq:final power}), the
setup of the expendable components $\underline{s}^{opt}$ that minimizes
(\ref{eq:cost function EMS}) can be found by maximizing $\mathcal{P}\left(\theta_{refl},\varphi_{refl},\underline{s}\right)$,
since $\mathcal{A}\left(\theta,\varphi\right)$ is not affected by
$\underline{s}$. This is yielded by applying the phase conjugation
principle \cite{Oliveri 2023b} to obtain the following closed form
solution\begin{equation}
s_{pq}^{opt}=\arg\left\{ \min_{s_{pq}=\left\{ 0,1\right\} }\left[\xi_{pq}\left(\theta_{refl},\varphi_{refl}\right)-\angle\Gamma^{TM}\left(\underline{g};s_{pq};\mathbf{k}_{inc}\right)\right]\right\} \label{eq:EMS design}\end{equation}
($p=1,...,P$; $q=1,...,Q$) where $\angle\cdot$ stands for the phase
term and\begin{equation}
\xi_{pq}\left(\theta_{refl},\varphi_{refl}\right)=\angle E_{inc}^{TM}\left(\mathbf{r}_{pq}\right)+k\left(x_{p}\sin\theta_{refl}\cos\varphi_{refl}+y_{q}\sin\theta_{refl}\sin\varphi_{refl}\right)\label{eq:}\end{equation}
is the \emph{ideal} phase compensation profile \cite{Oliveri 2023b}.
This means that for each $pq$-th ($p=1,...,P$; $q=1,...,Q$) meta-atom,
the optimal setup of the unit-cell status ($s_{pq}=0$ or $s_{pq}=1$)
is that which minimizes the mismatch between the actual local reflection
phase {[}i.e., $\angle\Gamma^{TM}\left(\underline{g};s_{pq};\mathbf{k}_{inc}\right)${]}
and the {}``ideal'' one {[}i.e., $\xi_{pq}\left(\theta_{refl},\varphi_{refl}\right)${]}.

\noindent Thanks to the closed-form design process (\ref{eq:EMS design})
and the possibility to pre-compute the entries of $\overline{\overline{\Gamma}}$
(e.g., through full-wave numerical simulations \cite{Oliveri 2022b}),
the synthesis of $\underline{s}^{opt}$ to enable a desired \emph{EMS}
functionality turns out extremely efficient even when large apertures
are at hand, the computational complexity growing linearly with the
number $P\times Q$ of meta-atoms of the \emph{EMS}. Moreover, such
a design framework may be generalized to account for multiple expendable
components in each unit cell to enable, as an example, a finer control
of the local reflection coefficients.

\section{\noindent Numerical and Experimental Validation\label{sec:Results}}

\noindent The aim of this section is twofold. On the one hand, it
is devoted to show the design as well as the results from the numerical
assessment of an \emph{OTP} meta-atom synthesized according to the
procedure in Sect. \ref{sub:Meta-Atom-Design-(MAD)} by taking into
account the non-ideal response of the class of expendable components
at hand. On the other hand, it is concerned with the numerical and
the experimental proofs of the effectiveness of an \emph{OTP-EMS},
based on the designed \emph{OTP} meta-atom, to fulfill the requirements
of a typical \emph{SEME} scenario.

\noindent Towards this end, a unitary ($E_{inc}^{TM}=1$ {[}V/m{]})
time-harmonic plane wave operating at $f_{0}=5.5$ {[}GHz{]} has been
used to illuminate an \emph{EMS} built on a $5.1\times10^{-4}$ {[}m{]}-thick
Rogers 4350B substrate with dielectric relative permittivity and dielectric
loss tangent equal to $\varepsilon_{r}=3.66$ and $\tan\delta=4.0\times10^{-3}$,
respectively, a standard $35$ {[}$\mu$m{]} copper thickness being
used for all metallizations.

\subsection{\emph{OTP} Meta-Atom Design\label{sub:ODR-Meta-Atom-Design}}

\noindent To design an effective \emph{OTP} meta-atom, it is mandatory
to have an accurate numerical model of the expendable component to
faithfully predict its behaviour when it is either on or off. Indeed,
non-idealities in the \emph{RF} response, often neglected in the device
data-sheets, can severely impact the arising local reflection tensor
with non-negligible deviations from its nominal value. Towards this
end, a set of $50$ copies of the chosen fuse, which is suitable for
printed circuit boards with surface mount technology (namely, the
LittleFuse R451) has been characterized. More specifically, each fuse
has been mounted on a \emph{SMA} connector {[}Fig. 1(\emph{b}){]}
and the resulting \emph{RF} properties have been measured at $f_{0}$
{[}Fig. 1(\emph{c}){]} to derive the average resistance/inductance
values reported in Tab. I, which have been then used for the full-wave
simulation of the \emph{OTP} meta-atom behavior. As for the reference
meta-atom layout, the geometry in \cite{Yang 2017}\cite{Oliveri 2022}
has been selected. It consists of a square patch with two edges short-circuited
to the ground plane through two surface-mounted fuses and two vias
{[}Fig. 1(\emph{a}){]}. By injecting a sufficient current at the center
of the patch, both fuses can be simultaneously broken so that the
arising structure features a \emph{single}-bit-per-atom reconfigurability.
The values of the geometrical descriptors in Fig. 1(\emph{a}) have
been optimized at $f_{0}$ according to the procedure in Sect. \ref{sub:Meta-Atom-Design-(MAD)}
($\underline{g}^{opt}$ - Tab. II) to yield the full-wave Ansys HFSS
\cite{HFSS 2021} simulated reflection magnitude {[}Fig. 3(\emph{a}){]}
and phase difference between the $s=1$ and $s=0$ states {[}Fig.
3(\emph{b}){]} when the illumination comes from broadside (i.e., $\left(\theta_{inc},\varphi_{inc}\right)=\left(0,0\right)$
{[}deg{]}). The plots in Fig. 3 show that, despite the non-ideal fuse
response, the optimized \emph{OTP} unit cell gives a reflection efficiency
above $50\%$ for both polarizations {[}i.e., $\left|\Gamma^{\zeta}\left(\underline{g}^{opt};s;\mathbf{k}_{inc}\right)\right|>-3$
{[}dB{]} ($\zeta=\left\{ TE,TM\right\} $) when $f=f_{0}$ - Fig.
3(\emph{a}){]} as well as a phase difference $\Delta\Gamma^{\zeta}$,
which is only $20\%$ far from the ideal $180$ {[}deg{]} one {[}i.e.,
$\Delta\Gamma^{\zeta}\approx145$ {[}deg{]} , $\zeta=\left\{ TE,TM\right\} $,
at $f=f_{0}$ - Fig. 3(\emph{b}){]} and perfectly identical for the
two polarizations {[}Fig. 3(\emph{b}){]}.

\subsection{\emph{OTP}-\emph{EMS}s Design and Full-Wave Validation\label{sub:ODR-EMSs-Design-and}}

\noindent This section is devoted to prove that an \emph{OTP-EMS},
implemented as a regular arrangement of $P\times Q$ identical \emph{OTP}
meta-atoms, is able to afford the user-required anomalous reflection
by configuring the status $\underline{s}$ of its expendable components
according to Sect. \ref{sub:OTP-EMS-Synthesis}.

\noindent The first illustrative example refers to the design of an
\emph{EMS} of $P\times Q=30\times30$ \emph{OTP} meta-atoms {[}Fig.
2(\emph{a}){]} that, when illuminated from broadside, reflects the
beam towards the angular direction $\left(\theta_{refl},\varphi_{refl}\right)=\left(-30,0\right)$
{[}deg{]}. Figure 4(\emph{a}) shows the map of the ideal phase profile
$\xi_{pq}\left(\theta_{refl},\varphi_{refl}\right)$ ($p=1,...,P$;
$q=1,...,Q$), while Figure 4(\emph{b}) reports the values of the
phase of the local reflection coefficient, $\angle\Gamma^{TM}\left(\underline{g};s_{pq};\mathbf{k}_{inc}\right)$
($p=1,...,P$; $q=1,...,Q$), obtained by setting the expendable components
according to (\ref{eq:EMS design}) {[}Fig. 4(\emph{c}){]}. As it
can be noticed {[}Fig. 4(\emph{a}) vs. Fig. 4(\emph{b}){]}, there
is a phase mismatch between the two distributions owing to the limited
control of the local reflection coefficient since each meta-atom has
only two states {[}Fig. 4(\emph{c}){]}. Such a deviation from the
target phase profile causes the presence of a quantization lobe in
the corresponding pattern with a magnitude similar to that of the
main beam as shown in Fig. 5 ($\varphi=0$ pattern cut). 

\noindent The presence of unwanted quantization lobes is a well-known
limitation of 1-bit \emph{EMS}s \cite{Yang 2017b}-\cite{Shekhawat 2025},
but despite the potential beam control losses with respect to the
use of more complex meta-atoms\cite{Oliveri 2024b}, most existing
designs focus on 1-bit phase control because they considerably simplify
the fabrication, while providing a good trade-off between design complexity
and performance \cite{Shekhawat 2025}\cite{Tang 2023}. Otherwise,
such \emph{undesired} lobes may be mitigated/avoided by increasing
the number of expendable components per unit cell, hence reducing
the phase quantization error, as shown by the pattern generated by
an \emph{ideal} unit cell (i.e., $\angle\Gamma^{TM}\left(\underline{g};s_{pq};\mathbf{k}_{inc}\right)=\xi_{pq}\left(\theta_{refl},\varphi_{refl}\right)$;
$p=1,...,P$; $q=1,...,Q$) in Fig. 5. Nevertheless, it is worthwhile
to point out that, despite the simple structure and the arising sub-optimal
reflection properties (Fig. 3), the conceived proof-of-concept \emph{EMS}
arrangement reflects the impinging beam towards the desired anomalous
direction (Fig. 5) by simply modifying the status of the expendable
components of a subset of the whole number of \emph{EMS} cells {[}Fig.
4(\emph{c}){]}. This is a first proof of the practical feasibility
of the \emph{OTP-EMS} concept as a strategy for the implementation
of low-cost, mass-production oriented, and customizable \emph{EM}
wave manipulation devices.

\noindent The next experiment is aimed at analyzing the dependence
of the pattern control capabilities on the size of the \emph{EMS}
aperture. Figure 6 shows the plots of $\mathcal{F}\left(\theta,\varphi;\underline{g}^{opt},\underline{s}^{opt}\right)$
in the $\varphi=0$ {[}deg{]} plane when setting $\left(\theta_{refl},\varphi_{refl}\right)=\left(-30,0\right)$
{[}deg{]} and varying the number of $P\times Q$ identical \emph{OTP}
unit cells of the square (i.e., $P=Q$) \emph{EMS}. Regardless its
area $\Omega$, the \emph{EMS} always implements the required anomalous
beam focusing and, as expected, the beamwidth reduces when the number
of meta-atoms per side $P$ is increased (Fig. 6). Vice-versa, the
amount of reflected energy is significantly reduced when the aperture
is very limited (e.g., $P=10$ - grey line in Fig. 6) owing to the
passive nature of the structure at hand.

\noindent To assess the robustness of the \emph{OTP-EMS} design principle
to the variation of the incidence angle $\left(\theta_{inc},\varphi_{inc}\right)$,
a set of $P\times Q=30\times30$ \emph{EMS} layouts has been synthesized
in the third test case by changing the illumination direction $\theta_{inc}$
, while setting $\theta_{refl}\triangleq\theta_{inc}-30$ {[}deg{]}
($\varphi_{inc}=\varphi_{refl}=0$ {[}deg{]}). Although the \emph{OTP}
meta-atom has not been designed specifically to deal with incidences
different from broadside, the arising pattern $\mathcal{F}\left(\theta,0;\underline{g}^{opt},\underline{s}^{opt}\right)$
always fulfils the reflection target (Fig. 7). 

\noindent The \emph{EMS} feature of yielding different reflection
directions in correspondence with the same impinging excitation is
assessed next. Towards this end, a set of $P\times Q=30\times30$
\emph{OTP-EMS} has been configured to reflect a beam coming from $\left(\theta_{inc},\varphi_{inc}\right)=\left(0,0\right)$
{[}deg{]} towards the variable direction $\left(\theta_{refl},0\right)$,
$\theta_{refl}\in\left\{ -10,-20,-30,-40\right\} $ {[}deg{]}. The
reflection patterns in Fig. 8 suggest the following takeaways: (\emph{i})
the designer can easily adjust the reflection angle within a wide
range just paying a {}``cos''-type scan degradation; (\emph{ii})
the position of the major secondary lobe is always predictable since
it is symmetric to the Snell's reflection direction.

\noindent The last example of the {}``\emph{Numerical Validation}''
deals with a full-wave analysis, carried out with Ansys HFSS \cite{HFSS 2021},
of the behaviour of \emph{OTP-EMS} when accounting for the non-idealities
of the finite arrangements at hand (e.g., non-periodic mutual coupling
among the meta-atoms, higher-order surface modes excited within the
substrate, edge diffraction and truncation effects, etc...). More
specifically, different \emph{EMS} apertures, $P\times Q=\left\{ 15\times15,20\times20\right\} $,
have been modeled with Ansys HFSS \cite{HFSS 2021} by using the finite-element
boundary-integral (\emph{FEBI}) method to avoid any periodic boundary
approximation. The comparison between the analytic and the numerically-computed
reflected patterns when $\left(\theta_{inc},\varphi_{inc}\right)=\left(0,0\right)$
{[}deg{]} and $\left(\theta_{refl},\varphi_{inc}\right)=\left(-30,0\right)$
{[}deg{]} (Fig. 9) proves that the \emph{OTP-EMS} design is robust
to the non-idealities of the structure as well as the reliability
of the synthesis guidelines.

\noindent For the sake of completeness, the corresponding 3D plots
in the $uv$-plane of the reflected patterns are reported in Fig.
10 to further highlight the agreement between the analytical prediction
and the full-wave computed one.

\subsection{Experimental Validation\label{sub:ODR-EMSs-Experimental}}

\noindent Finally, the \emph{OTP-EMS} concept has been experimentally
validated to demonstrate its practical feasibility and its potentialities
in terms of wave manipulation capabilities as well as realization
modularity (Fig. 11). Accordingly, a $P\times Q=10\times10$ layout
($\approx25$ {[}cm{]}-side panel) has been realized on a Rogers 4350B
substrate with a standard \emph{PCB} prototyping processes {[}Figs.
11(\emph{a})-11(\emph{b}){]}. For the sake of measurement simplicity
and without loss of generality, a single polarization illumination
has been considered. More specifically, the transmitting device and
the receiving one were a linearly-polarized log-periodic antenna and
a wideband horn antenna, respectively, while the reflected field has
been measured in a standard semi-anechoic chamber setup.

\noindent The matching between measured and simulated patterns, normalized
to the transmitting/receiving antenna gains, in Fig. 11(\emph{c})
confirms also experimentally the wave manipulation functionality of
the \emph{OTP-EMS} prototype in Figs. 11(\emph{a})-11(\emph{b}), despite
this latter has been fabricated with a low cost HW by avoiding advanced
electronic circuitry. Moreover, it is worth to point out that the
fully-passive layout of \emph{OTP-EMS} operates as a static reflector
once it has been hardware-coded, thus there are neither power consumption,
nor network overhead, nor protocol adjustment once deployed.

\section{\noindent Conclusions\label{sec:Conclusions-and-Remarks}}

\noindent The concept of \emph{OTP-EMS} has been introduced to address
the request of simple, inexpensive, and mass-production oriented solutions
for flexible \emph{EM} wave manipulations in \emph{SEME}s. By integrating
\emph{expendable components} at the atomic level, \emph{OTP-EMS}s
allow the designer to simultaneously achieve passive-static and no-maintenance
operations, while also yielding modular fabrication and configurable
reflection properties. To validate such a concept, an \emph{OTP} meta-atom
structure, which supports a desired wave manipulation feature, has
been first synthesized. Successively, the effectiveness and the flexibility
of the resulting \emph{EMS}s have been analytically, numerically,
and experimentally assessed.

\noindent From both the numerical analysis and the experimental validation,
the following main outcomes can be drawn:

\begin{itemize}
\item \noindent the synthesized \emph{OTP} meta-atom \emph{}(Fig. 3) has
a satisfactory dual-polarization response despite the simplicity of
its layout and the presence of non-idealities caused by the off-the-shelf
expendable components (Fig. 1);
\item \emph{OTP-EMS}s support flexible \emph{EM} wave manipulation capabilities
with satisfactory performance when varying both incidence (Fig. 7)
and reflection directions (Fig. 8) without requiring very large apertures
(Fig. 6);
\item the performance of \emph{}the \emph{OTP-EMS} prototype confirms the
full-wave numerical prediction (Fig. 11) as well as they assess the
reliability of the \emph{OTP} technology in fulfilling user-defined
reflection requirements.
\end{itemize}
\noindent Future works, beyond the scope of this manuscript, but still
in this research framework, will be aimed at extending the \emph{OTP-EMS}
performance to other frequency bands as well as at introducing more
complex unit cells layouts to address other/multiple requirements/constraints.

\section*{\noindent Acknowledgements}

\noindent This work benefited from the networking activities carried
out within the Project {}``ICSC National Centre for HPC, Big Data
and Quantum Computing (CN HPC)'' funded by the European Union - NextGenerationEU
within the PNRR Program (CUP: E63C22000970007), the Project {}``Smart
ElectroMagnetic Environment in TrentiNo - SEME@TN'' funded by the
Autonomous Province of Trento (CUP: C63C22000720003), the Project
DICAM-EXC (Grant L232/2016) funded by the Italian Ministry of Education,
Universities and Research (MUR) within the {}``Departments of Excellence
2023-2027'' Program (CUP: E63C22003880001), the Project {}``INSIDE-NEXT
- Indoor Smart Illuminator for Device Energization and Next-Generation
Communications'' funded by the Italian Ministry for Universities
and Research within the PRIN 2022 Program (CUP: E53D23000990001),
the Project {}``AURORA - Smart Materials for Ubiquitous Energy Harvesting,
Storage, and Delivery in Next Generation Sustainable Environments''
funded by the Italian Ministry for Universities and Research within
the PRIN-PNRR 2022 Program, and the Structural Project 6GWINET - Project
RESTART (Partnership on {}``Telecommunications of the Future'' -
PE00000001) funded by European Union under the Italian National Recovery
and Resilience Plan (NRRP) of NextGenerationEU. A. Massa wishes to
thank E. Vico for her never-ending inspiration, support, guidance,
and help.

\newpage
\section*{FIGURE CAPTIONS}

\begin{itemize}
\item \textbf{Figure 1.} \emph{Problem Scenario} - Sketch of (\emph{a})
the \emph{OTP} meta-atom layout and photos of (\emph{b}) the expendable
component and (\emph{c}) the measurement setup.
\item \textbf{Figure 2.} \emph{EMS Geometry} ($P\times Q=30\times30$) -
Sketch of (\emph{a}) the numerical model of the \emph{OTP} \emph{EMS}
along with (\emph{b}) a detail of the meta-atom layout with both intact
and burnt expendable components.
\item \textbf{Figure 3.} \emph{OTP Meta-Atom Design} ($\left(\theta_{inc},\varphi_{inc}\right)=\left(0,0\right)$
{[}deg{]}) - Behaviour of the $\zeta$-th ($\zeta=\left\{ TE,TM\right\} $)
component of (\emph{a}) $\left|\Gamma^{\zeta}\left(\underline{g}^{opt};s;\mathbf{k}_{inc}\right)\right|$
and (\emph{b}) $\Delta\Gamma^{\zeta}$ versus the frequency.
\item \textbf{Figure 4.} \emph{OTP-EMS Synthesis} ($P\times Q=30\times30$,
$\left(\theta_{inc},\varphi_{inc}\right)=\left(0,0\right)$ {[}deg{]},
$\left(\theta_{refl},\varphi_{inc}\right)=\left(-30,0\right)$ {[}deg{]})
- Plots of (\emph{a}) the ideal phase profile, \{$\xi_{pq}\left(\theta_{refl},\varphi_{refl}\right)$;
$p=1,...,P$; $q=1,...,Q$\}, and (\emph{b}) the phase distribution
of the local reflection coefficient, \{$\angle\Gamma^{TM}\left(\underline{g}^{opt};s_{pq};\mathbf{k}_{inc}\right)$;
$p=1,...,P$; $q=1,...,Q$\}, afforded by (\emph{c}) the $\underline{s}^{opt}$-configured
\emph{OTP-EMS}.
\item \textbf{Figure 5.} \emph{OTP-EMS Synthesis} ($P\times Q=30\times30$,
$\left(\theta_{inc},\varphi_{inc}\right)=\left(0,0\right)$ {[}deg{]},
$\left(\theta_{refl},\varphi_{inc}\right)=\left(-30,0\right)$ {[}deg{]})
- Plot of the $\varphi=0$ {[}deg{]} cut of the pattern $\mathcal{F}\left(\theta,\varphi;\underline{g},\underline{s}\right)$
reflected by an ideal \emph{EMS} and the synthesized \emph{OTP-EMS}
($\underline{g}=\underline{g}^{opt}$, $\underline{s}=\underline{s}^{opt}$).
\item \textbf{Figure 6.} \emph{OTP-EMS Synthesis} ($\left(\theta_{inc},\varphi_{inc}\right)=\left(0,0\right)$
{[}deg{]}, $\left(\theta_{refl},\varphi_{inc}\right)=\left(-30,0\right)$
{[}deg{]}) - Plot of the $\varphi=0$ {[}deg{]} cut of the reflected
pattern $\mathcal{F}\left(\theta,\varphi;\underline{g}^{opt},\underline{s}^{opt}\right)$
generated by \emph{OTP-EMS} panels with different sizes (i.e., number
of $P\times Q$ meta-atoms).
\item \textbf{Figure 7.} \emph{OTP-EMS Synthesis} ($P\times Q=30\times30$,
$\varphi_{inc}=\varphi_{refl}=0$ {[}deg{]}) - Plots of the $\varphi=0$
{[}deg{]} cut of the reflected pattern $\mathcal{F}\left(\theta,\varphi;\underline{g}^{opt},\underline{s}^{opt}\right)$
synthesized when $\left(\theta_{inc},\theta_{refl}\right)=\left(0,-30\right)$
{[}deg{]}, $\left(\theta_{inc},\theta_{refl}\right)=\left(10,-20\right)$
{[}deg{]}, and $\left(\theta_{inc},\theta_{refl}\right)=\left(20,-10\right)$
{[}deg{]}.
\item \textbf{Figure 8.} \emph{OTP-EMS Synthesis} ($P\times Q=30\times30$,
$\varphi_{inc}=\varphi_{refl}=0$ {[}deg{]}, $\theta_{inc}=0$ {[}deg{]})
- Plots of the $\varphi=0$ {[}deg{]} cut of the reflected pattern
$\mathcal{F}\left(\theta,\varphi;\underline{g}^{opt},\underline{s}^{opt}\right)$
synthesized when $\theta_{refl}\in\left\{ -10,-20,-30,-40\right\} $
{[}deg{]}.
\item \textbf{Figure 9.} \emph{OTP-EMS Numerical Validation} ($\left(\theta_{inc},\varphi_{inc}\right)=\left(0,0\right)$
{[}deg{]}, $\left(\theta_{refl},\varphi_{inc}\right)=\left(-30,0\right)$
{[}deg{]}) - Comparison between the analytically-computed reflected
pattern $\mathcal{F}\left(\theta,\varphi;\underline{g}^{opt},\underline{s}^{opt}\right)$
and the full-wave simulated one along the $\varphi=0$ {[}deg{]} cut
for an \emph{EMS} panel composed of (\emph{a}) $P\times Q=15\times15$
and (\emph{b}) $P\times Q=20\times20$ \emph{OTP} meta-atoms.
\item \textbf{Figure 10.} \emph{OTP-EMS Numerical Validation} ($\left(\theta_{inc},\varphi_{inc}\right)=\left(0,0\right)$
{[}deg{]}, $\left(\theta_{refl},\varphi_{inc}\right)=\left(-30,0\right)$
{[}deg{]}) - Color-maps in the $u-v$ plane of (\emph{a})(\emph{c})
the analytically-computed and (\emph{b})(\emph{d}) the full-wave simulated
reflected pattern $\mathcal{F}\left(\theta,\varphi;\underline{g}^{opt},\underline{s}^{opt}\right)$
for an \emph{EMS} panel composed of (\emph{a})(\emph{b}) $P\times Q=15\times15$
and (\emph{c})(\emph{d}) $P\times Q=20\times20$ \emph{OTP} meta-atoms.
\item \textbf{Figure 11.} \emph{OTP-EMS Experimental Validation} ($P\times Q=10\times10$,
$\left(\theta_{inc},\varphi_{inc}\right)=\left(0,0\right)$ {[}deg{]},
$\left(\theta_{refl},\varphi_{inc}\right)=\left(-30,0\right)$ {[}deg{]})
- Pictures of (\emph{a}) the complete \emph{OTP-EMS} prototype \emph{}and
(\emph{b}) a detail of its meta-atom layout along with (\emph{c})
the comparison between the analytically-computed reflected pattern
$\mathcal{F}\left(\theta,0;\underline{g}^{opt},\underline{s}^{opt}\right)$
and the measured one in the $\varphi=0$ {[}deg{]} cut.
\end{itemize}

\section*{TABLE CAPTIONS}

\begin{itemize}
\item \textbf{Table I.} \emph{OTP Meta-Atom Design -} Measured electrical
properties of the selected expendable component LittleFuse R451.
\item \textbf{Table II.} \emph{OTP Meta-Atom Design -} Values of the optimized
geometrical descriptors, $\underline{g}^{opt}$.
\end{itemize}
\newpage
\begin{center}~\end{center}

\begin{center}\begin{tabular}{c}
\includegraphics[%
  width=0.75\columnwidth]{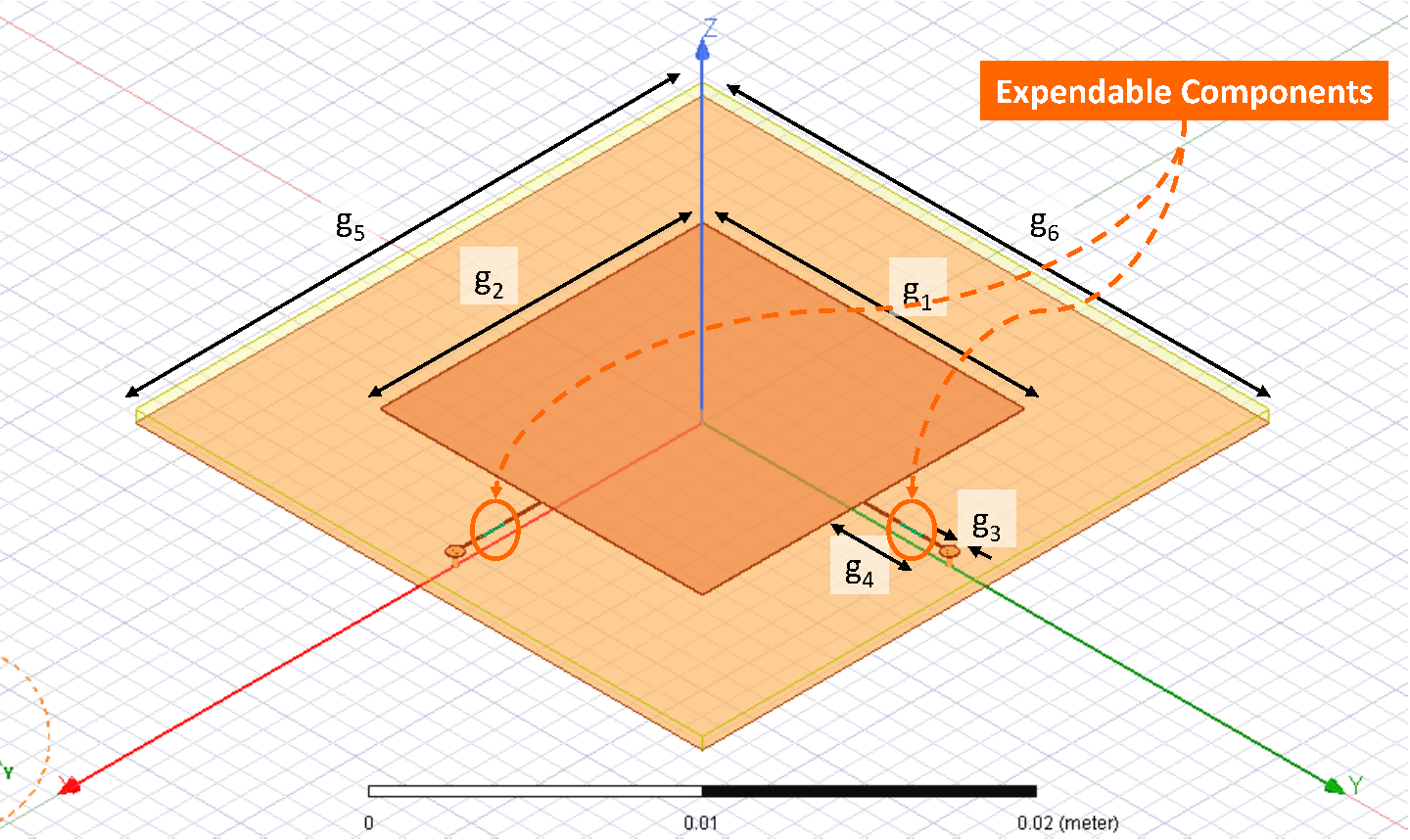}\tabularnewline
(\emph{a})\tabularnewline
\includegraphics[%
  width=0.70\columnwidth,
  keepaspectratio]{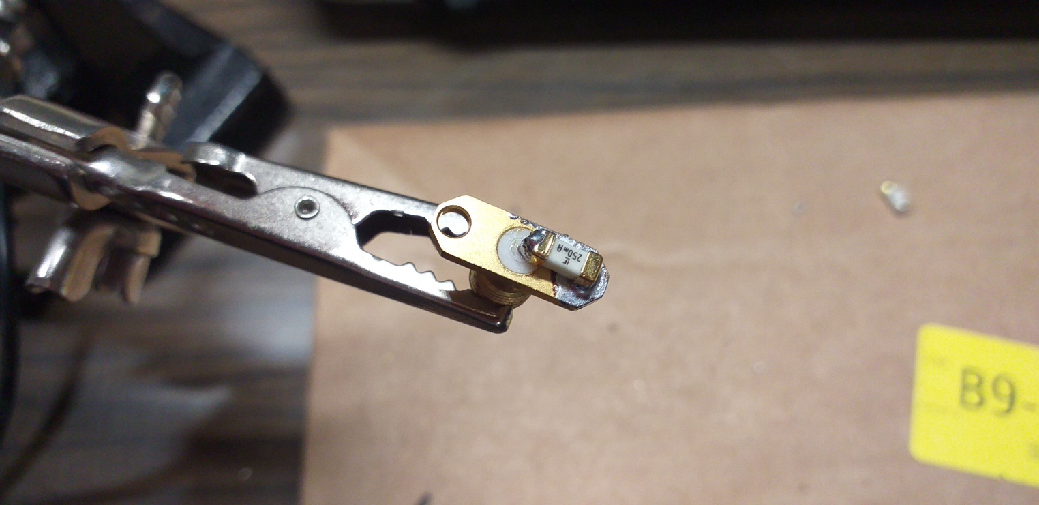}\tabularnewline
(\emph{b})\tabularnewline
\includegraphics[%
  width=0.70\columnwidth,
  keepaspectratio]{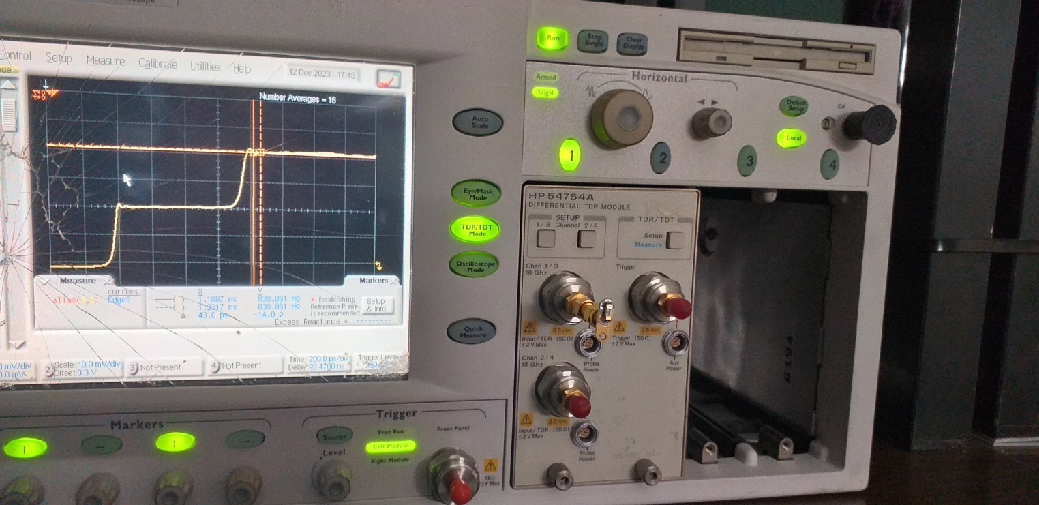}\tabularnewline
(\emph{c})\tabularnewline
\end{tabular}\end{center}

\begin{center}~\vfill\end{center}

\begin{center}\textbf{Fig. 1 - G. Oliveri} \textbf{\emph{et al.,}}
{}``One-Time Programmable Passive Electromagnetic Skins''\end{center}

\newpage
\begin{center}~\end{center}

\begin{center}\vfill\end{center}

\begin{center}\begin{tabular}{c}
\includegraphics[%
  width=0.90\columnwidth]{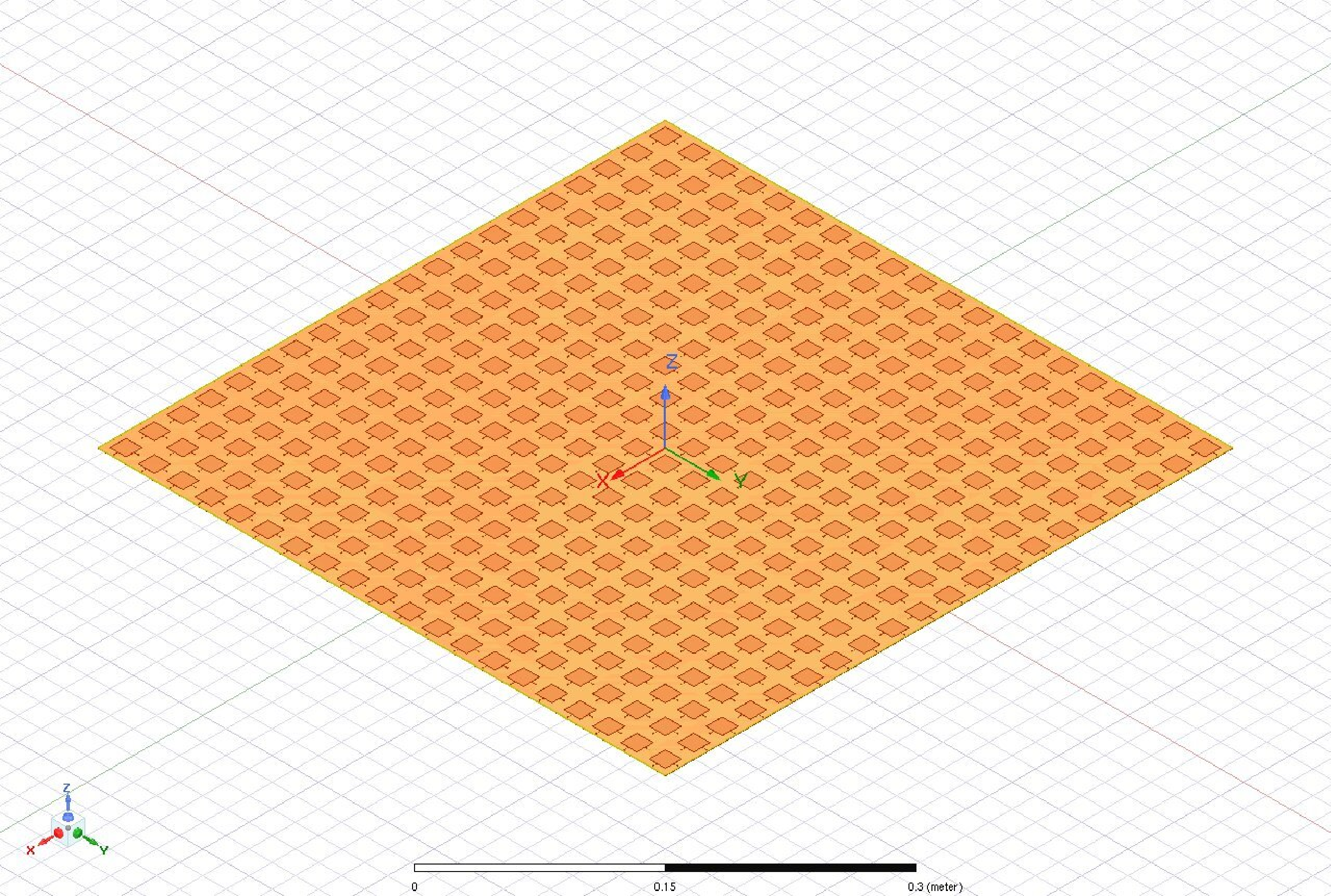}\tabularnewline
(\emph{a})\tabularnewline
\includegraphics[%
  width=0.90\columnwidth]{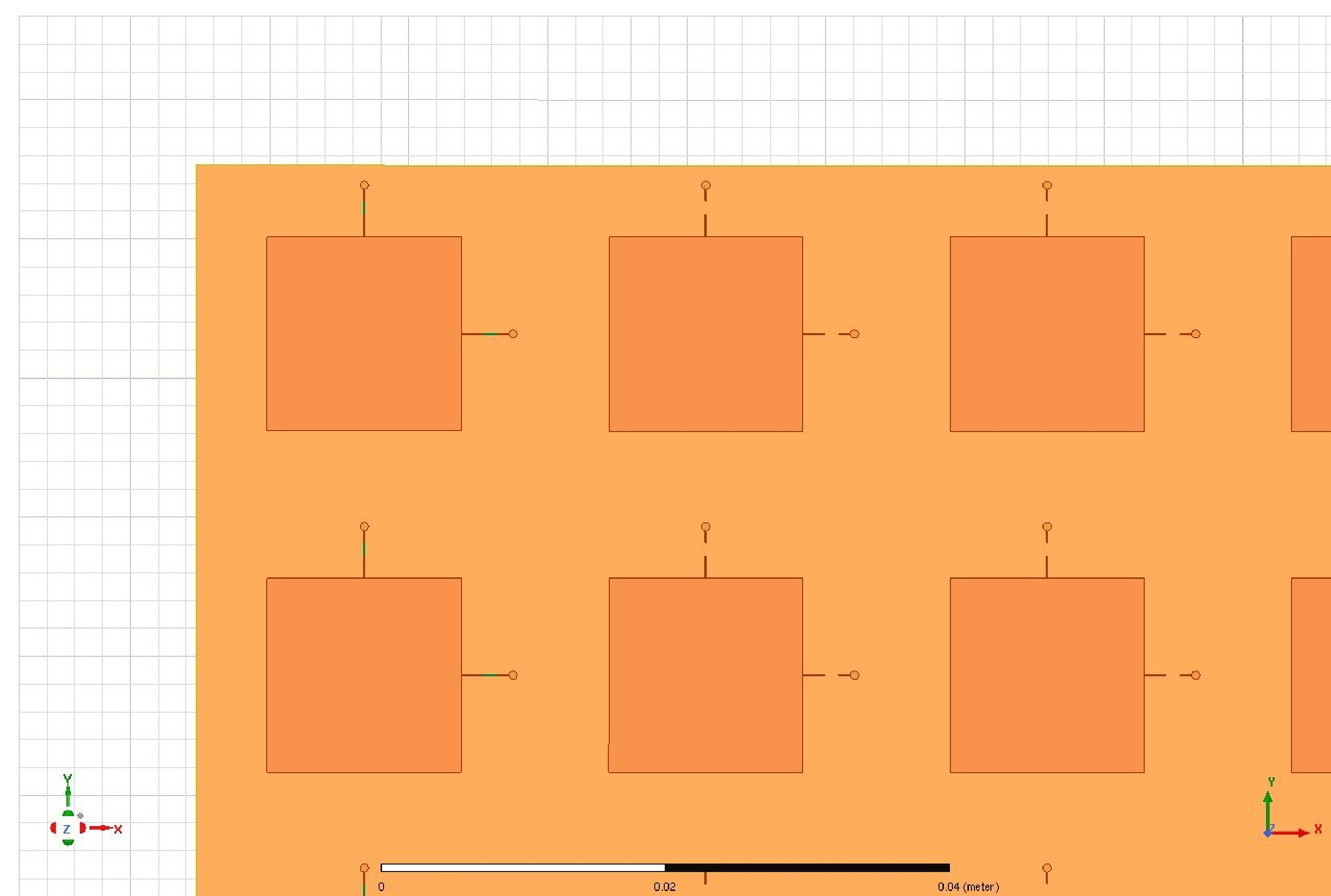}\tabularnewline
(\emph{b})\tabularnewline
\end{tabular}\end{center}

\begin{center}~\vfill\end{center}

\begin{center}\textbf{Fig. 2 - G. Oliveri} \textbf{\emph{et al.,}}
{}``One-Time Programmable Passive Electromagnetic Skins''\end{center}

\newpage
\begin{center}~\end{center}

\begin{center}\begin{tabular}{c}
\includegraphics[%
  width=0.85\columnwidth]{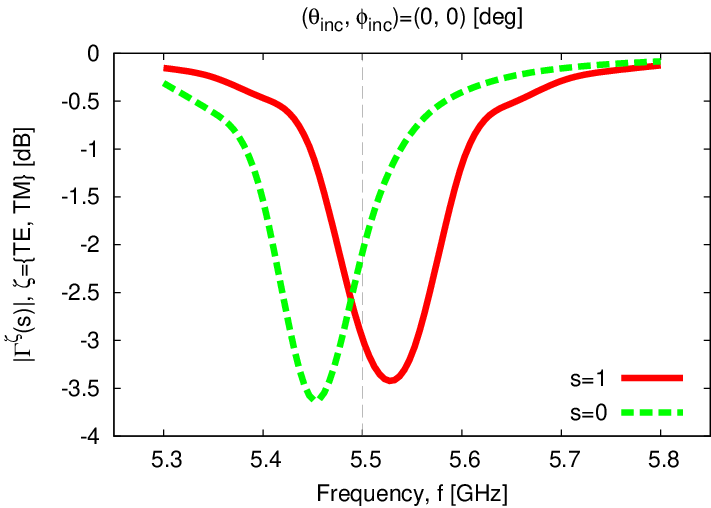}\tabularnewline
(\emph{a})\tabularnewline
\includegraphics[%
  width=0.85\columnwidth]{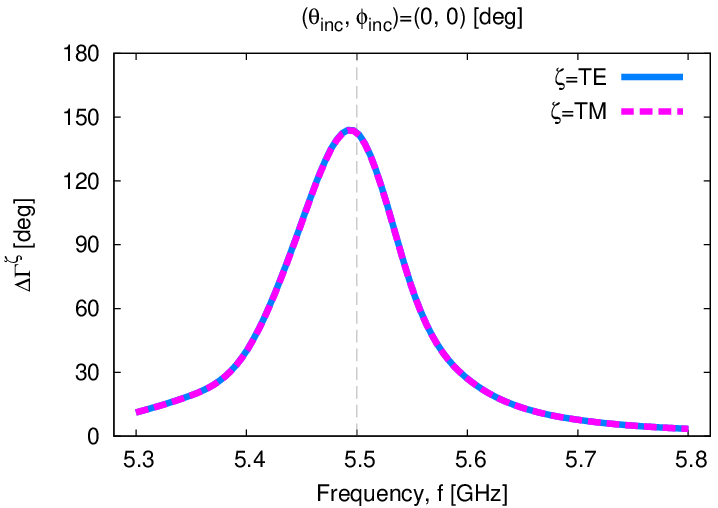}\tabularnewline
(\emph{b})\tabularnewline
\end{tabular}\end{center}

\begin{center}\vfill~\end{center}

\begin{center}\textbf{Fig. 3 - G. Oliveri} \textbf{\emph{et al.,}}
{}``One-Time Programmable Passive Electromagnetic Skins''\end{center}

\newpage
\begin{center}~\end{center}

\begin{center}\begin{tabular}{c}
\includegraphics[%
  width=0.47\columnwidth]{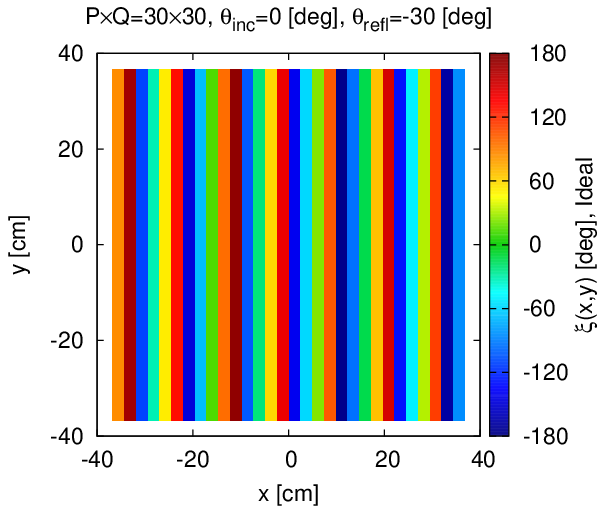}\tabularnewline
(\emph{a})\tabularnewline
\includegraphics[%
  width=0.47\columnwidth]{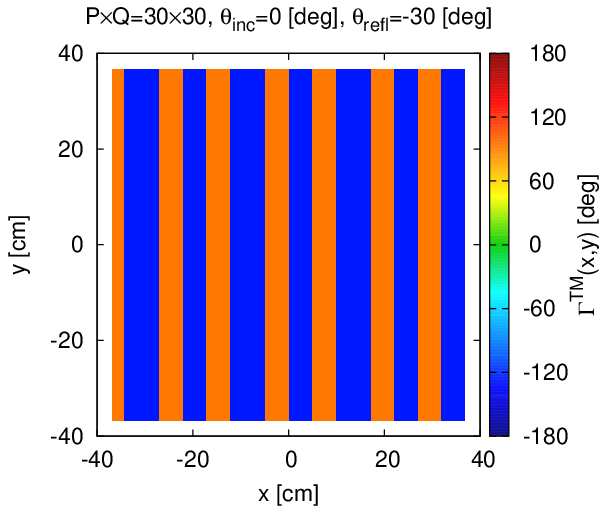}\tabularnewline
(\emph{b})\tabularnewline
\includegraphics[%
  width=0.47\columnwidth]{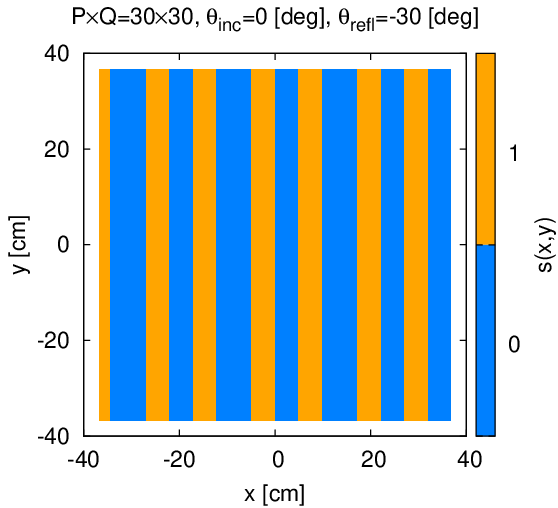}\tabularnewline
(\emph{c})\tabularnewline
\end{tabular}\end{center}

\begin{center}\textbf{Fig. 4 - G. Oliveri} \textbf{\emph{et al.,}}
{}``One-Time Programmable Passive Electromagnetic Skins''\end{center}

\newpage
\begin{center}~\end{center}

\begin{center}\vfill\end{center}

\begin{center}\includegraphics[%
  width=0.95\columnwidth]{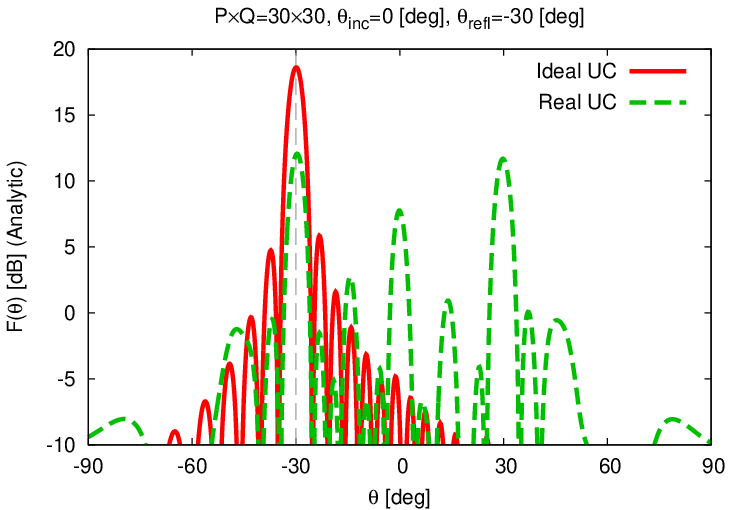}\end{center}

\begin{center}~\vfill\end{center}

\begin{center}\textbf{Fig. 5 - G. Oliveri} \textbf{\emph{et al.,}}
{}``One-Time Programmable Passive Electromagnetic Skins''\end{center}

\newpage
\begin{center}~\end{center}

\begin{center}\vfill\end{center}

\begin{center}\includegraphics[%
  width=0.95\columnwidth]{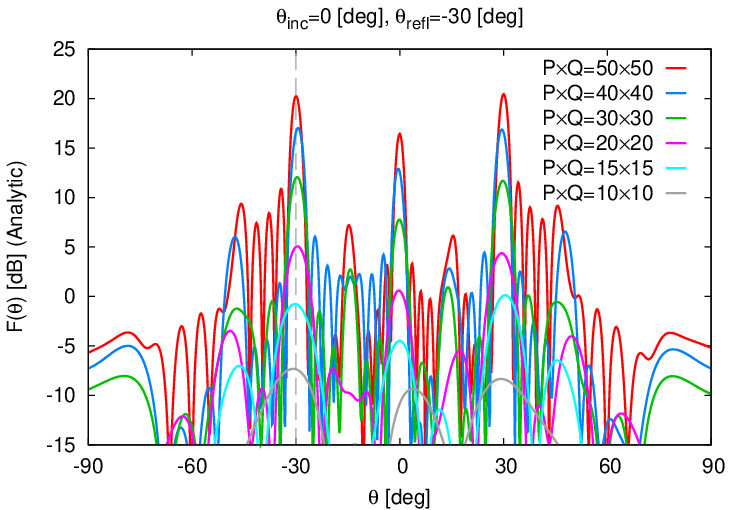}\end{center}

\begin{center}~\vfill\end{center}

\begin{center}\textbf{Fig. 6 - G. Oliveri} \textbf{\emph{et al.,}}
{}``One-Time Programmable Passive Electromagnetic Skins''\end{center}

\newpage
\begin{center}~\end{center}

\begin{center}\vfill\end{center}

\begin{center}\includegraphics[%
  width=0.95\columnwidth]{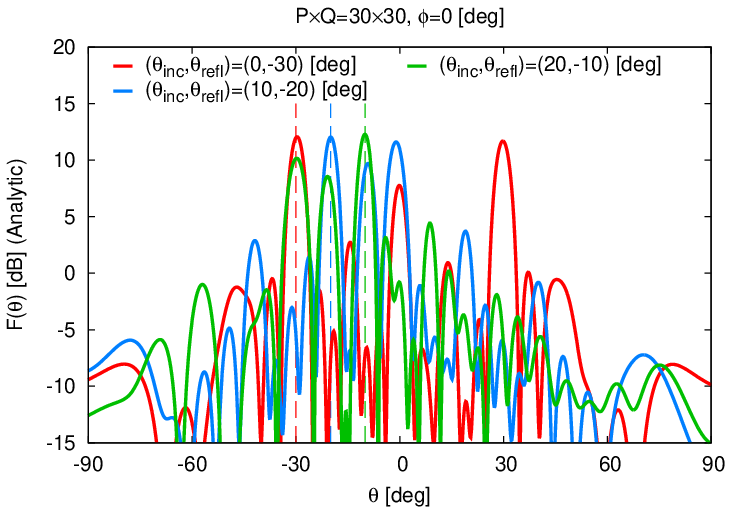}\end{center}

\begin{center}~\vfill\end{center}

\begin{center}\textbf{Fig. 7 - G. Oliveri} \textbf{\emph{et al.,}}
{}``One-Time Programmable Passive Electromagnetic Skins''\end{center}

\newpage
\begin{center}~\end{center}

\begin{center}\vfill\end{center}

\begin{center}\includegraphics[%
  width=0.95\columnwidth]{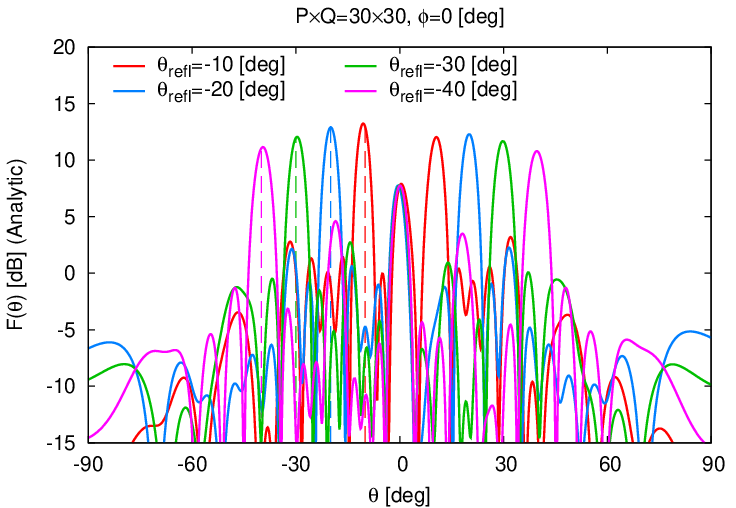}\end{center}

\begin{center}~\vfill\end{center}

\begin{center}\textbf{Fig. 8 - G. Oliveri} \textbf{\emph{et al.,}}
{}``One-Time Programmable Passive Electromagnetic Skins''\end{center}

\newpage
\begin{center}~\end{center}

\begin{center}\vfill\end{center}

\begin{center}\begin{tabular}{c}
\includegraphics[%
  width=0.80\textwidth]{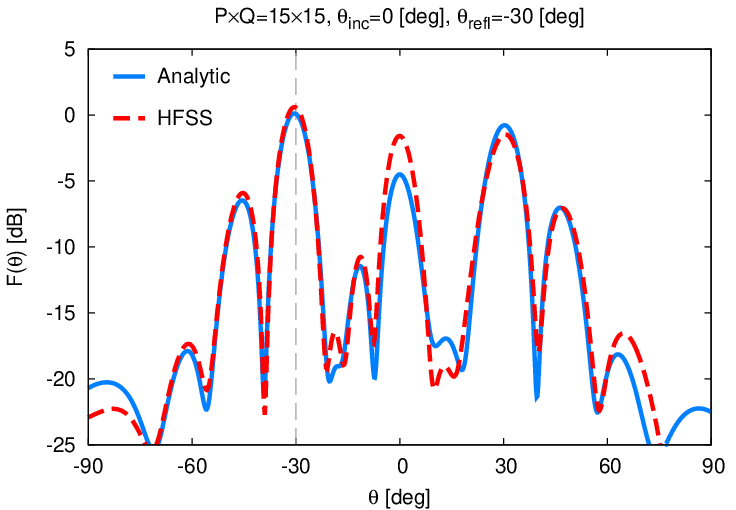}\tabularnewline
(\emph{a})\tabularnewline
\includegraphics[%
  width=0.80\textwidth]{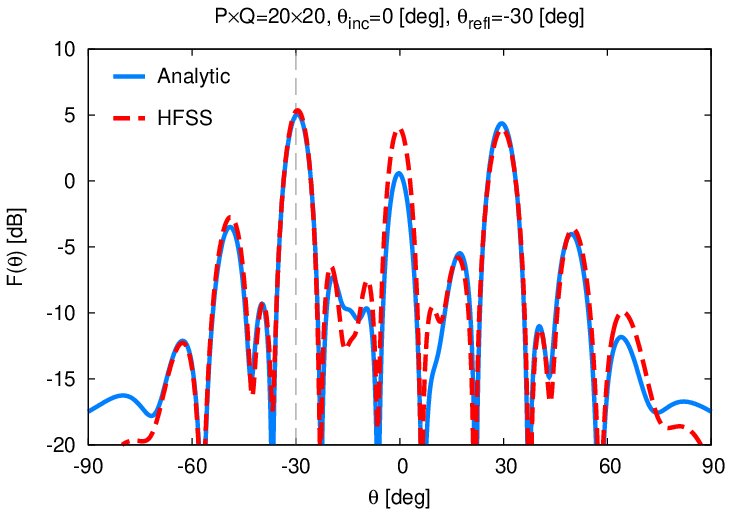}\tabularnewline
(\emph{b})\tabularnewline
\end{tabular}\end{center}

\begin{center}~\vfill\end{center}

\begin{center}\textbf{Fig. 9 - G. Oliveri} \textbf{\emph{et al.,}}
{}``One-Time Programmable Passive Electromagnetic Skins''\end{center}

\newpage
\begin{center}~\end{center}

\begin{center}\vfill\end{center}

\begin{center}\begin{tabular}{ccc}
&
\emph{Analytic}&
\emph{HFSS}\tabularnewline
\begin{sideways}
~~~~~~~~~~~~~~~~~~~~~~~~~~~~~~$15\times15$%
\end{sideways}&
\includegraphics[%
  width=0.45\columnwidth]{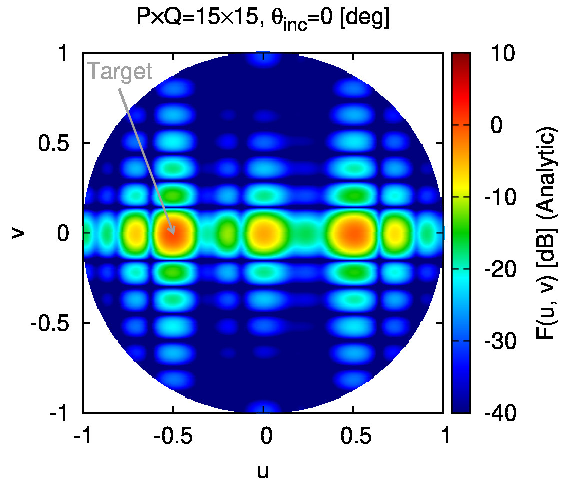}&
\includegraphics[%
  width=0.45\columnwidth]{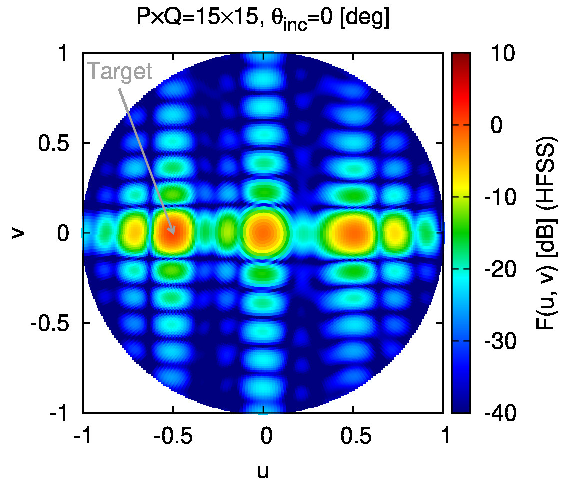}\tabularnewline
&
(\emph{a})&
(\emph{b})\tabularnewline
\begin{sideways}
~~~~~~~~~~~~~~~~~~~~~~~~~~~~~~$20\times20$%
\end{sideways}&
\includegraphics[%
  width=0.45\columnwidth]{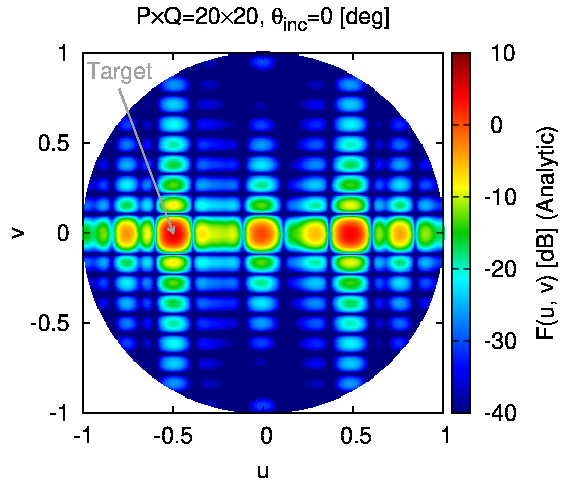}&
\includegraphics[%
  width=0.45\columnwidth]{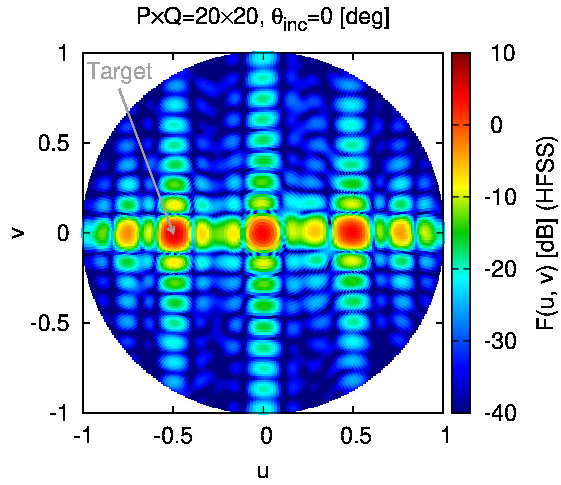}\tabularnewline
&
(\emph{c})&
(\emph{d})\tabularnewline
\end{tabular}\end{center}

\begin{center}~\vfill\end{center}

\begin{center}\textbf{Fig. 10 - G. Oliveri} \textbf{\emph{et al.,}}
{}``One-Time Programmable Passive Electromagnetic Skins''\end{center}

\newpage
\begin{center}~\vfill\end{center}

\begin{center}\begin{tabular}{cc}
\includegraphics[%
  width=0.50\columnwidth,
  keepaspectratio]{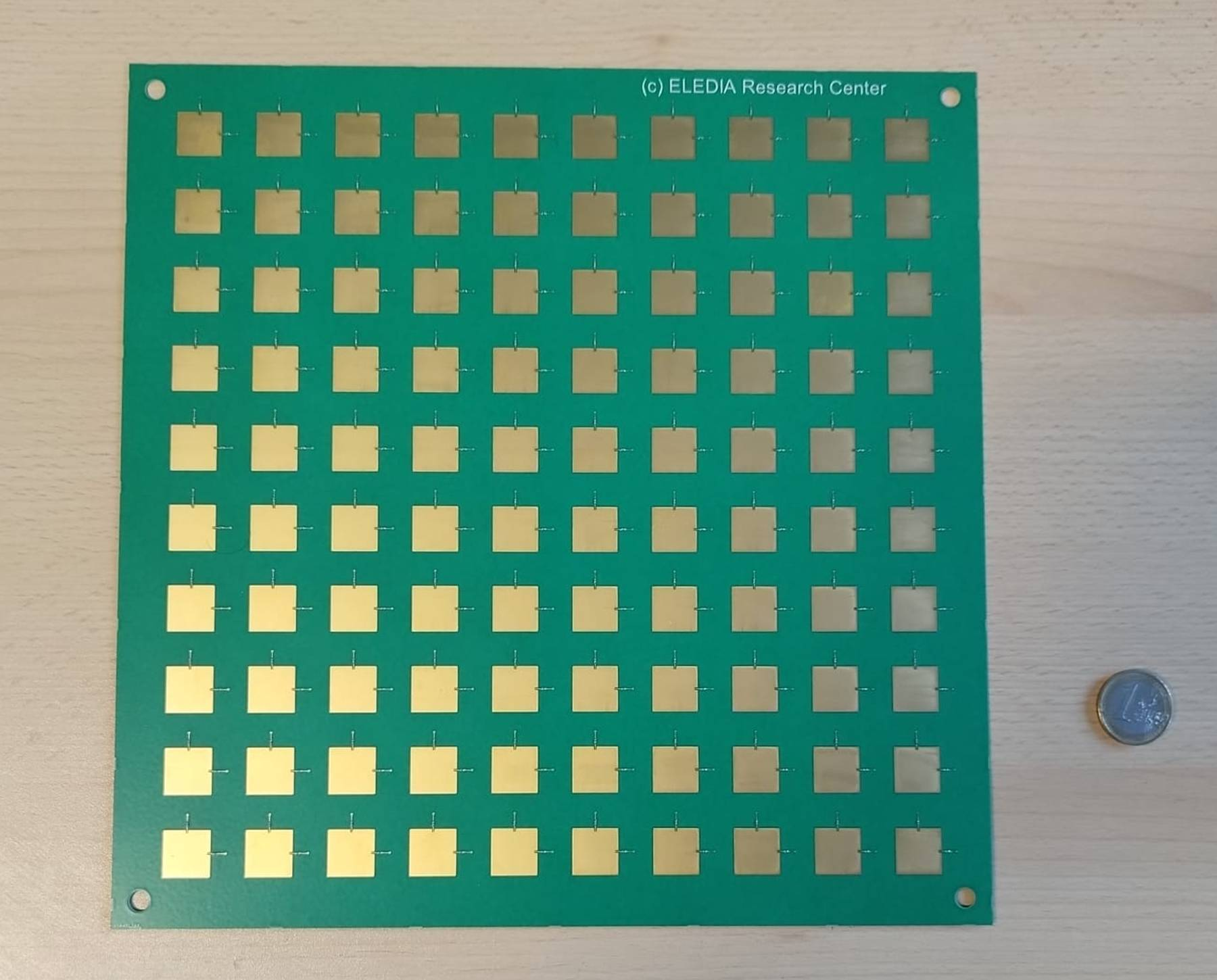}&
\includegraphics[%
  width=0.30\columnwidth,
  keepaspectratio]{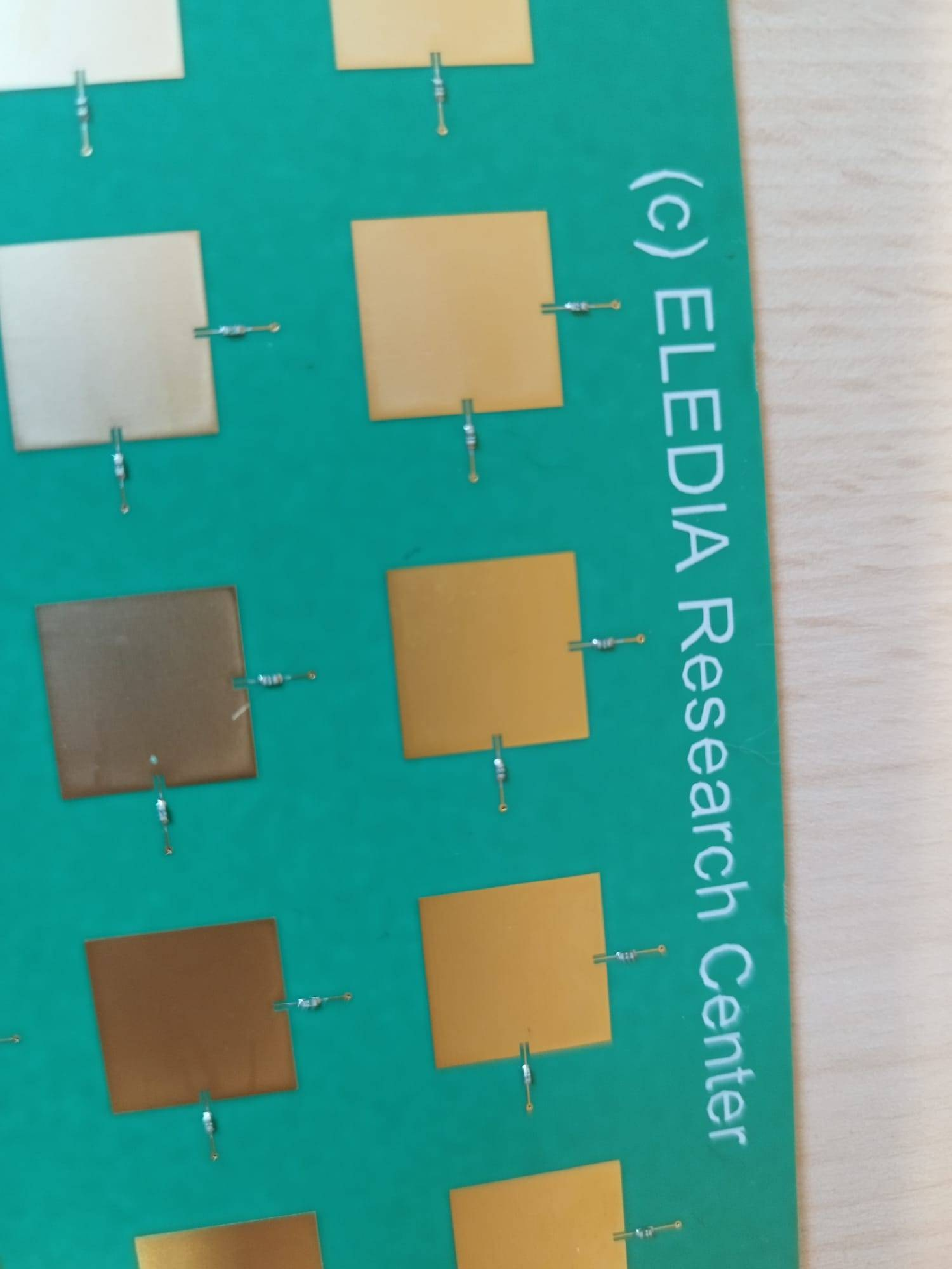}\tabularnewline
(\emph{a})&
(\emph{b})\tabularnewline
\multicolumn{2}{c}{}\tabularnewline
\multicolumn{2}{c}{\includegraphics[%
  width=0.90\textwidth]{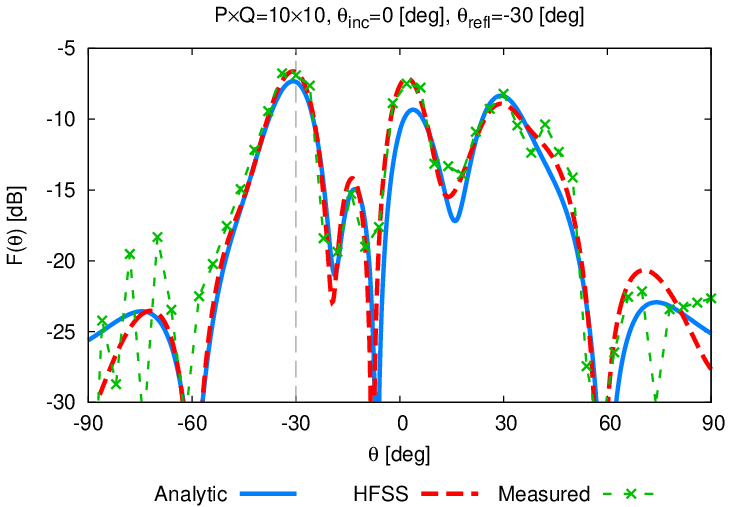}}\tabularnewline
\multicolumn{2}{c}{(\emph{c})}\tabularnewline
\end{tabular}\end{center}

\begin{center}~\vfill\end{center}

\begin{center}\textbf{Fig. 11 - G. Oliveri} \textbf{\emph{et al.,}}
{}``One-Time Programmable Passive Electromagnetic Skins''\end{center}

\newpage
\begin{center}~\vfill\end{center}

\begin{center}\begin{tabular}{|c|c|c|}
\hline 
&
\textbf{Average Value}&
\textbf{Unit}\tabularnewline
\hline
\hline 
Intact Fuse Resistance&
$0.6$&
{[}ohm{]}\tabularnewline
\hline 
Intact Fuse Inductance&
$3.0$&
{[}nH{]}\tabularnewline
\hline 
Broken Fuse Resistance&
$<0.1$&
{[}ohm{]}\tabularnewline
\hline 
Broken Fuse Inductance&
$<0.1$&
{[}nH{]}\tabularnewline
\hline
\end{tabular}\end{center}

\begin{center}~\vfill\end{center}

\begin{center}\textbf{Tab. I - G. Oliveri} \textbf{\emph{et al.}}\textbf{,}
{}``One-Time Programmable Passive Electromagnetic Skins''\end{center}

\newpage
\begin{center}~\vfill\end{center}

\begin{center}\begin{tabular}{|c|c|c|}
\hline 
\textbf{Descriptor}&
\textbf{Value}&
\textbf{Unit}\tabularnewline
\hline
\hline 
Square patch edge, $g_{1}=g_{2}$&
$13.95$&
{[}mm{]}\tabularnewline
\hline 
Pin radius, $g_{3}$&
$0.3$&
{[}mm{]}\tabularnewline
\hline 
Microstrip length, $g_{4}$&
$3.71$&
{[}mm{]}\tabularnewline
\hline 
Unit cell spacing, $g_{5}=g_{6}$&
$0.45$&
$\lambda@f_{0}$\tabularnewline
&
$2.45$&
{[}cm{]}\tabularnewline
\hline
\end{tabular}\end{center}

\begin{center}~\vfill\end{center}

\begin{center}\textbf{Tab. II - G. Oliveri} \textbf{\emph{et al.}}\textbf{,}
{}``One-Time Programmable Passive Electromagnetic Skins''\end{center}\newpage

\end{document}